\documentclass[twocolumn]{aastex62}

\usepackage{multirow}
\usepackage{booktabs}
\usepackage{mathtools}

\definecolor{red}{rgb}{1.0,0.0,0.0}
\definecolor{blue}{rgb}{0.0,0.5,1.0}

\received{Sep.\ 8, 2019}
\revised{Jan.\ 27, 2020}
\accepted{Feb.\ 18, 2020}
\submitjournal{AJ}

\shorttitle{The HOSTS survey results}
\shortauthors{Ertel et al.}

\begin{document}

\title{The HOSTS survey for exozodiacal dust: Observational results from the complete survey}

\correspondingauthor{Steve Ertel}
\email{sertel@email.arizona.edu}

\author{Ertel,~S.}
\affiliation{Large Binocular Telescope Observatory, 933 North Cherry Avenue, Tucson, AZ 85721, USA}
\affiliation{Steward Observatory, Department of Astronomy, University of Arizona, 993 N. Cherry Ave, Tucson, AZ, 85721,
USA}

\author{Defr\`ere,~D.}
\affiliation{Space sciences, Technologies \& Astrophysics Research (STAR) Institute, University of Li\`ege, Li\`ege,
Belgium}

\author{Hinz,~P.}
\affiliation{Steward Observatory, Department of Astronomy, University of Arizona, 993 N. Cherry Ave, Tucson, AZ, 85721,
USA}

\author{Mennesson,~B.}
\affiliation{Jet Propulsion Laboratory, California Institute of Technology, 4800 Oak Grove Dr., Pasadena, CA 91109,
USA}

\author{Kennedy,~G.~M.}
\affiliation{Department of Physics, University of Warwick, Gibbet Hill Road, Coventry CV4 7AL, UK}

\author{Danchi,~W.~C.}
\affiliation{NASA Goddard Space Flight Center, Exoplanets \& Stellar Astrophysics Laboratory, Code 667, Greenbelt, MD
20771, USA}

\author{Gelino,~C.}
\affiliation{Jet Propulsion Laboratory, California Institute of Technology, 4800 Oak Grove Dr., Pasadena, CA 91109,
USA}

\author{Hill,~J.~M.}
\affiliation{Large Binocular Telescope Observatory, University of Arizona, 933 N. Cherry Avenue, Tucson, AZ 85721, USA}

\author{Hoffmann,~W.~F.}
\affiliation{Steward Observatory, Department of Astronomy, University of Arizona, 993 N. Cherry Ave, Tucson, AZ, 85721,
USA}

\author{Mazoyer,~J.}
\affiliation{Jet Propulsion Laboratory, California Institute of Technology, 4800 Oak Grove Dr., Pasadena, CA 91109,
USA}

\author{Rieke,~G.}
\affiliation{Steward Observatory, Department of Astronomy, University of Arizona, 993 N. Cherry Ave, Tucson, AZ, 85721,
USA}

\author{Shannon,~A.}
\affiliation{LESIA, Observatoire de Paris, PSL Research University, CNRS, Sorbonne Universit\'e, Universit\'e Paris Diderot, Sorbonne Paris
  Cit\'e, 5 Place Jules Janssen, 92195 Meudon, France}
\affiliation{Department of Astronomy and Astrophysics, The Pennsylvania State University, State College, PA 16801, USA}
\affiliation{Center for Exoplanets and Habitable Worlds, The Pennsylvania State University, State College, PA 16802,
USA}

\author{Stapelfeldt,~K.}
\affiliation{Jet Propulsion Laboratory, California Institute of Technology, 4800 Oak Grove Dr., Pasadena, CA 91109,
USA}

\author{Spalding,~E.}
\affiliation{Steward Observatory, Department of Astronomy, University of Arizona, 993 N. Cherry Ave, Tucson, AZ, 85721,
USA}

\author{Stone,~J.~M.}
\altaffiliation{Hubble fellow.}
\affiliation{Steward Observatory, Department of Astronomy, University of Arizona, 993 N. Cherry Ave, Tucson, AZ, 85721,
USA}

\author{Vaz,~A.}
\affiliation{Steward Observatory, Department of Astronomy, University of Arizona, 993 N. Cherry Ave, Tucson, AZ, 85721,
USA}

\author{Weinberger,~A.~J.}
\affiliation{Department of Terrestrial Magnetism, Carnegie Institution of Washington, 5241 Broad Branch Road NW,
Washington, DC, 20015, USA}

\author{Willems,~P.}
\affiliation{Jet Propulsion Laboratory, California Institute of Technology, 4800 Oak Grove Dr., Pasadena, CA 91109,
USA}

\author{Absil,~O.}
\altaffiliation{F.R.S.-FNRS Research Associate}
\affiliation{Space sciences, Technologies \& Astrophysics Research (STAR) Institute, University of Li\`ege, Li\`ege,
Belgium}

\author{Arbo,~P.}
\affiliation{Steward Observatory, Department of Astronomy, University of Arizona, 993 N. Cherry Ave, Tucson, AZ, 85721,
USA}

\author{Bailey,~V.~P.}
\affiliation{Jet Propulsion Laboratory, California Institute of Technology, 4800 Oak Grove Dr., Pasadena, CA 91109,
USA}

\author{Beichman,~C.}
\affiliation{NASA Exoplanet Science Institute, MS 100-22, California Institute of Technology, Pasadena, CA 91125, USA}

\author{Bryden,~G.}
\affiliation{Jet Propulsion Laboratory, California Institute of Technology, 4800 Oak Grove Dr., Pasadena, CA 91109,
USA}

\author{Downey,~E.~C.}
\affiliation{Steward Observatory, Department of Astronomy, University of Arizona, 993 N. Cherry Ave, Tucson, AZ, 85721,
USA}

\author{Durney,~O.}
\affiliation{Steward Observatory, Department of Astronomy, University of Arizona, 993 N. Cherry Ave, Tucson, AZ, 85721,
USA}

\author{Esposito,~S.}
\affiliation{INAF-Osservatorio Astrofisico di Arcetri, Largo E. Fermi 5, I-50125 Firenze, Italy}

\author{Gaspar,~A.}
\affiliation{Steward Observatory, Department of Astronomy, University of Arizona, 993 N. Cherry Ave, Tucson, AZ, 85721,
USA}

\author{Grenz,~P.}
\affiliation{Steward Observatory, Department of Astronomy, University of Arizona, 993 N. Cherry Ave, Tucson, AZ, 85721,
USA}

\author{Haniff,~C.~A.}
\affiliation{Cavendish Laboratory, University of Cambridge, JJ Thomson Avenue, Cambridge CB3 0HE, UK}

\author{Leisenring,~J.~M.}
\affiliation{Steward Observatory, Department of Astronomy, University of Arizona, 993 N. Cherry Ave, Tucson, AZ, 85721,
USA}

\author{Marion,~L.}
\affiliation{Space sciences, Technologies \& Astrophysics Research (STAR) Institute, University of Li\`ege, Li\`ege,
Belgium}

\author{McMahon,~T.~J.}
\affiliation{Steward Observatory, Department of Astronomy, University of Arizona, 993 N. Cherry Ave, Tucson, AZ, 85721,
USA}

\author{Millan-Gabet,~R.}
\affiliation{NASA Exoplanet Science Institute, MS 100-22, California Institute of Technology, Pasadena, CA 91125, USA}

\author{Montoya,~M.}
\affiliation{Steward Observatory, Department of Astronomy, University of Arizona, 993 N. Cherry Ave, Tucson, AZ, 85721,
USA}

\author{Morzinski,~K.~M.}
\affiliation{Steward Observatory, Department of Astronomy, University of Arizona, 993 N. Cherry Ave, Tucson, AZ, 85721,
USA}

\author{Perera,~S.}
\affiliation{Max Planck Institut f\"ur Astronomie, K\"onigstuhl 17, 69117 Heidelberg, Germany}
\affiliation{Steward Observatory, Department of Astronomy, University of Arizona, 993 N. Cherry Ave, Tucson, AZ, 85721,
USA}

\author{Pinna,~E.}
\affiliation{INAF-Osservatorio Astrofisico di Arcetri, Largo E. Fermi 5, I-50125 Firenze, Italy}

\author{Pott,~J.-U.}
\affiliation{Max Planck Institut f\"ur Astronomie, K\"onigstuhl 17, 69117 Heidelberg, Germany}

\author{Power,~J.}
\affiliation{Large Binocular Telescope Observatory, University of Arizona, 933 N. Cherry Avenue, Tucson, AZ 85721, USA}

\author{Puglisi,~A.}
\affiliation{INAF-Osservatorio Astrofisico di Arcetri, Largo E. Fermi 5, I-50125 Firenze, Italy}

\author{Roberge,~A.}
\affiliation{NASA Goddard Space Flight Center, Exoplanets \& Stellar Astrophysics Laboratory, Code 667, Greenbelt, MD
20771, USA}

\author{Serabyn,~E.}
\affiliation{Jet Propulsion Laboratory, California Institute of Technology, 4800 Oak Grove Dr., Pasadena, CA 91109,
USA}

\author{Skemer,~A.~J.}
\affiliation{Astronomy Department, University of California Santa Cruz, 1156 High Street, Santa Cruz, CA 95064, USA}

\author{Su,~K.~Y.~L.}
\affiliation{Steward Observatory, Department of Astronomy, University of Arizona, 993 N. Cherry Ave, Tucson, AZ, 85721,
USA}

\author{Vaitheeswaran,~V.}
\affiliation{Steward Observatory, Department of Astronomy, University of Arizona, 993 N. Cherry Ave, Tucson, AZ, 85721,
USA}

\author{Wyatt,~M.~C.}
\affiliation{Institute of Astronomy, University of Cambridge, Madingley Road, Cambridge CB3 0HA, UK}



\begin{abstract}

The Large Binocular Telescope Interferometer (LBTI) enables nulling interferometric observations across the $N$~band (8~to 13\,$\mu$m) to
suppress a star's bright light and probe for faint circumstellar emission.  We present and statistically analyze the results from the
LBTI/HOSTS (Hunt for Observable Signatures of Terrestrial Systems) survey for exozodiacal dust.  By comparing our measurements to model
predictions based on the Solar zodiacal dust in the $N$~band, we estimate a 1\,$\sigma$ median sensitivity of 23\,zodis for early type stars
and 48\,zodis for Sun-like stars, where
1\,zodi is the surface density of habitable zone (HZ) dust in the Solar system.  Of the 38 stars observed, 10 show significant excess.  A
clear correlation of our detections with the presence of cold dust in the systems was found, but none with the stellar spectral type or age. 
The majority of Sun-like stars have relatively low HZ dust levels (best-fit median: 3\,zodis, 1\,$\sigma$ upper limit: 9\,zodis, 95\%
confidence: 27\,zodis based on our $N$~band measurements), while $\sim$20\% are significantly more dusty.  The Solar system's HZ dust content
is consistent with being typical.  Our median HZ dust level would not be a major limitation to the direct imaging search for Earth-like
exoplanets, but more precise constraints are still required, in particular to evaluate the impact of exozodiacal dust for the spectroscopic
characterization of imaged exo-Earth candidates.

\end{abstract}

\keywords{Exozodiacal dust (500), Debris disks (363), Habitable zone (696), Habitable planets (695)}


\section{Introduction}

Imaging habitable exoplanets (exo-Earth imaging) is one of the major challenges of modern astronomy.  The main technical challenges are the
required high contrast and small inner working angle resulting from the faintness of the planets and their proximity to the bright host
stars.  In addition, exozodiacal dust constitutes an \textit{astrophysical} challenge for exo-Earth imaging to be understood and potentially
to be overcome \citep{roberge2012}.  This analog to the zodiacal dust in our Solar system \citep{kelsall1998, dermott2002, nesvorny2010} is
expected to be present in and near the habitable zones (HZs) of the exo-Earth imaging mission target stars.  The presence of large amounts of
exozodiacal dust in a system represents a major source of photon noise that may render a faint planet undetectable.  Furthermore, spatial
structures in the dust distribution may add confusion and be misinterpreted as planets due to the limited angular resolution and signal-to
noise ratio of the observations \citep{defrere2012b}.  Smaller amounts of smoothly distributed dust may still make an imaged planet's
spectroscopic characterization prohibitively time consuming.  As a consequence, the occurrence rate and typical brightness of massive
exozodiacal dust systems affect the yield of future exo-Earth imaging missions and are thus important factors for the mission design
(aperture size, mission duration, target selection, \citealt{defrere2010, stark2015, stark2016, stark2019}).

In addition, studying the dust distribution provides present day insight into the characteristics of HZs around nearby stars
\citep{kral2017}.  Dust in and near the HZ of a star (HZ dust) has a temperature around 300\,K and is best detected near the peak of its
spectral energy distribution near 10$\,\mu$m.  This dust is distinct from colder dust in a debris disk further out in the system that is
typically detected photometrically in the far-infrared (exo-Kuiper belts) and can most often be explained by continuous dust production in an
equilibrium collisional cascade \citep{dohnanyi1969, backman1993}.  Inside these outer belts, many systems have dust at temperatures similar
to those in the asteroidal zone of the solar system \citep{morales2011, kennedy2014}, which may similarly originate from a local equilibrium
collisional cascade or have an origin similar or related to that of the HZ dust discussed below (including belt formation due to planet-disk
interaction, e.g., \citealt{ertel2012b, shannon2015}).  The HZ dust is also different from the hot excesses detected around nearby stars
using optical long baseline interferometry and usually attributed to dust emission even closer in \citep{absil2006, absil2013, ertel2014b},
while the mechanisms producing this hot dust may or may not be related to those producing the HZ dust \citep{kennedy2015b, rieke2016,
faramaz2017, kimura2018, sezestre2019}.

The HZ dust may be produced through collisions of planetesimals in an outer, Kuiper or asteroid-belt-like debris disk and migrate inward due
to Poynting-Robertson (PR) drag and stellar wind drag \citep{reidemeister2011, wyatt2005}.  The amount of dust that reaches the HZ may
then be used to constrain the presence of planets between the outer reservoir and the HZ that prevent a fraction of the dust from migrating
\citep{moro-martin2003, bonsor2018}.  Alternatively, the dust may be produced by comets sublimating or otherwise disintegrating when they
reach the HZ from further out in the system \citep{nesvorny2010, faramaz2017, marino2017, sezestre2019}, which is thought to be the main
source of zodiacal dust in the Solar system \citep{nesvorny2010, shannon2015, poppe2019}.  Thus, observations of HZ dust have the potential
to put constraints on the cometary activity in the system, providing insights into the dynamics of the outer regions \citep{bonsor2012,
bonsor2014, faramaz2017, marino2017} and the environmental conditions of potential rocky planets (cometary bombardment, delivery of water;
\citealt{kral2018}).  Other scenarios such as a recent, catastrophic collision near the HZ \citep[e.g.,][]{lisse2012, bonsor2013, meng2014,
su2019} or local dust production in a massive belt of planetesimals near the HZ are likely less common, but the fact that systems dominated
by such processes exist has important implications for the architecture and evolution of HZs.  Their frequency is yet to be determined beyond
the most extreme cases, but the existence of rare bright ones implies a population of more common faint ones \citep{kennedy2013}.  While
spatial dust structures may hinder exo-Earth imaging, studying them may also reveal the presence of otherwise currently undetectable HZ
planets and help to determine their properties \citep{stark2008, ertel2012b, shannon2015}.

Detecting exozodiacal dust is challenging due to the small separation from its host star\footnote{A separation of 1\,au at a typical distance
of 10\,pc for nearby stars corresponds to 0.1$''$.} and the dust temperature of a few 100\,K which means it emits predominantly in the
mid-infrared where it is outshone by the star.  Photometric observations to detect the dust excess emission are limited to a sensitivity of
a few per cent of the stellar emission due to flux calibration uncertainties and limitations in predicting the stellar photospheric flux.
This limit is significantly higher than measured for all but the most extreme and rare excesses.  Spectroscopic observations may slightly
improve over this sensitivity if silicate emission features can be detected \citep{ballering2014}.  Detecting scattered light from dust very
close to the star in visible light aperture polarization measurements has been unsuccessful, which puts important constraints on the
properties and origin of the hot, near-infrared detected dust \citep{marshall2016}.  Interferometry is required to spatially resolve the
thermal dust emission in the infrared and thus disentangle it from the host star.  This has been done successfully for the hot dust using
optical long baseline interferometry in the near-infrared \citep{absil2006, absil2013, defrere2012a, ertel2014b, ertel2016, nunez2017}.  In
the mid infrared where HZ dust is the brightest, nulling interferometry \citep{bracewell1979, hinz1998, hinz2000} has been used to suppress
the bright, unresolved star light and detect the faint, extended dust emission \citep{stock2010, millan-gabet2011, mennesson2014,
ertel2018a}.

In this paper we present and statistically analyze the complete data set from the HOSTS (Hunt for Observable Signatures of Terrestrial
planetary Systems) survey.  We have observed a sample of 38 nearby stars using the nulling mode of the Large Binocular Telescope
Interferometer (LBTI, \citealt{hinz2016}).  Our observations probed for HZ dust around the target stars with approximately five times better
sensitivity than past observations.  Thus, they provide the strongest direct constraints on the HZ dust contents of a sample of nearby
planetary systems and the strongest statistical constraints for future exo-Earth imaging mission target stars.

We describe our observations and data reduction in Sect.~\ref{sect_obs}.  In Sect.~\ref{sect_res} we present our basic results.  We discuss
our data quality and detection criteria (Sect.~\ref{sect_data}), and describe the conversion of the astrophysical null measurements to dust
levels in units of `1\,zodi', i.e., multiples of the vertical optical depth of the Solar system's HZ dust (Sect.~\ref{sect_null-to-zodi}). 
A discussion of our results is presented in Sect.~\ref{sect_disc}.  We start with extracting and discussing basic detection statistics that
we correlate with other parameters of the observed targets, such as stellar spectral type, age, and the presence of known cold dust
(Sect.~\ref{sect_disc_corr}).  We discuss the prospects of more detailed studies of our detections based on our available data and follow-up
observations with the LBTI are discussed in Sect.~\ref{sect_disc_followup}.  In Sect.~\ref{sect_disc_sample} we describe a deeper
statistical analysis of our data that provides the strongest possible constraints on the typical zodi level around future exo-Earth imaging
targets (Sect.~\ref{sect_free_form}), discuss the implications of our results for future exo-Earth imaging missions
(Sect.~\ref{sect_imp_exoearths}), and outline a path forward to further improve the LBTI's sensitivity and provide even stronger constraints
from a revived HOSTS survey (Sect.~\ref{sect_sensitivity}).  Our conclusions are presented in Sect.~\ref{sect_conc}.

\section{Observations and data reduction}

\label{sect_obs}
The observations for the HOSTS survey were carried out with the Large Binocular Telescope Interferometer (LBTI, \citealt{hinz2016})
following the strategy outlined in detail by \citet{ertel2018a}.  We used nulling interferometry in the $N'$ filter ($\lambda_{\rm c}$ =
11.11\,$\mu$m, $\Delta\lambda$ = 2.6\,$\mu$m) to combine the light from the two 8.4m apertures of the LBT and to suppress the light from the
central star through destructive interference.  The total flux transmitted through the interferometric null was measured on our NOMIC
(Nulling-Optimized Mid-Infrared Camera, \citealt{hoffmann2014}) detector and compared to a photometric observation of the target star to
determine the null leak (fraction of light transmitted).  Nodding and aperture photometry were used to subtract the variable telescope and
sky background.  Each observation of a science target (SCI) was paired with an identical observation of a calibration star (CAL) to determine
the instrumental null leak (nulling transfer function, the instrumental response to a point source).  The difference between the total null
leak and the instrumental null leak is the astrophysical null $N_{\text{as}}$, i.e., the source flux transmitted through the instrument due
to spatially resolved emission.  Multiple such calibrated science observations were executed (typically two to four) and typically grouped in
sequences of CAL--SCI--SCI--CAL for observing efficiency.

Science targets were selected according to target observability and priority from the full HOSTS target list compiled by
\citet{weinberger2015}.  This list consists of nearby, bright (\textsl{N}~$>$~1\,Jy) main sequence stars without known close binary
companions (within 1.5$''$).  Because of their low luminosities, stars with late spectral types need to be close to pass our brightness
limit and are thus relatively rare in our sample.  The sample can be separated into early type stars (spectral types A to F5) for which our
observations are most sensitive and Sun-like stars (spectral types F6 to K8) which are preferred targets for future exo-Earth imaging
missions.  The observed stars are listed with their relevant properties and observing dates in Table~\ref{tab_stars}.  About half of the
stars selected by \citet{weinberger2015} have been observed; the observed stars are representative of the full list with no significant
additional biases other than target observability during the observing nights when nulling was possible (see below).

Calibrators were selected following \citet{mennesson2014} using the catalogs of \citet{borde2002} and \citet{merand2005}, supplemented by
stars from the Jean-Marie Mariotti Center Stellar Diameter Catalog and the SearchCal tool (both \citealt{chelli2016}) where necessary.
Multiple calibrators were selected for each science target so that the same calibrator was typically not used repeatedly for the same science
target in order to minimize systematic errors due to imperfect knowledge of the calibrator stars (potential binarity or circumstellar
emission, uncertain diameter).

Observations were carried out in queue mode together with a variety of other observing programs using the LBTI, including high-contrast
direct imaging \citep[e.g.,][]{stone2018} and integral field spectroscopic observations \citep[e.g.,][]{stone2018spie, briesemeister2019}.
This increased the pool of nights to choose from for the nulling observations which are very demanding in terms of weather conditions.  A
total of ten nights of observing time per observing semester was allocated for the HOSTS survey over the 2016B, to 2018A semesters (40 nights
total), of which typically three to four nights per semester were used successfully while the rest was largely lost due to unsuitable
weather conditions (during which we often executed other, less demanding projects from our observing queue).

Data reduction followed the strategy outlined by \citet{defrere2016} with minor updates as described by \citet{ertel2018a}.  After a basic
reduction of each frame (nod subtraction, bad pixel correction), aperture photometry was performed on each single frame.  Three different
photometric apertures were used to (1)~cover~one resolution element of the single aperture point spread function (PSF), to (2)~optimize the
photon and read noise limited signal-to-noise ratio for extended emission analogous to the Solar system zodiacal dust, and to (3)~include all
plausible extended $N$~band dust emission from the system.  These apertures were discussed and motivated in detail by \citet{ertel2018a}.
They respectively have radii of 8\,pix (143\,mas), 13\,pix (233\,mas), and the EEID\footnote{
$\text{EEID} = 1\,\text{au} \times \sqrt{L_\star / L_\odot}$, the Earth Equivalent Insolation Distance from the star at which a body
receives the same energy density as Earth does from the Sun.} plus one full width at half maximum (FWHM) of the single aperture PSF
($\text{EEID}+313$\,mas, `conservative aperture').  The raw null depths and their uncertainties were determined using the null self
calibration method (NSC, \citealt{mennesson2011, hanot2011, defrere2016, mennesson2016a}), combining all frames recorded within a given nod
for a statistical analysis.  These measurements within an observing sequence of a science target were then combined and the corresponding
calibrator observations were used to calibrate the null measurements.  These calibrated astrophysical null measurements for each aperture
and each science target are listed in Table~\ref{tab_measurements}.  All raw and calibrated HOSTS data are available to the public through
the LBTI Archive (http://lbti.ipac.caltech.edu/).

\section{Results}
\label{sect_res}

\begin{figure*}
 \centering
 \includegraphics[angle=0,width=\linewidth]{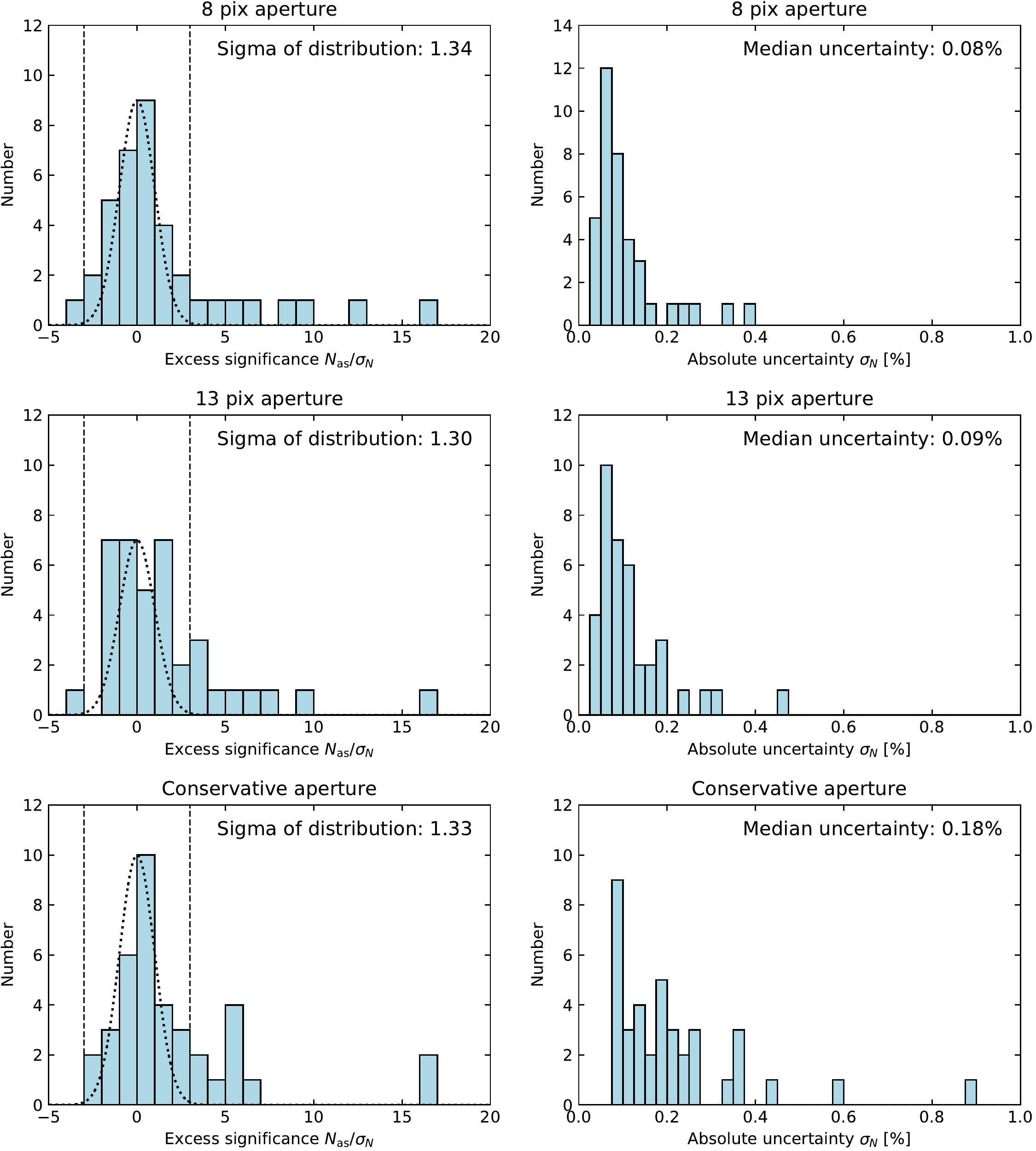}
 \caption{Histograms of astrophysical null measurements and uncertainties for all hosts survey observations.}
 \label{fig_hist_nulls}
\end{figure*}

\subsection{Data quality and detection criteria}
\label{sect_data}

The astrophysical null and zodi measurements derived from the HOSTS survey are listed in Table~\ref{tab_measurements}.
Fig.~\ref{fig_hist_nulls} shows the astrophysical null measurements and sensitivities reached for all stars and the three apertures used. 
The distributions of the significance $N_{\text{as}}/\sigma_N$ of the measurements (the ratio between the calibrated, astrophysical null
measurement $N_{\text{as}}$ and its measurement uncertainty $\sigma_N$) are generally well behaved, consistent with a Gaussian distribution
around a significance of $N_{\text{as}}/\sigma_N = 0$ and a tail of detections at $N_{\text{as}}/\sigma_N > 3$.  This can be expected for a
sample in which a fraction of stars have no detectable excesses, while the other stars do have significant excess.  The standard deviation
of the Gaussian component (measured for stars with $N_{\text{as}}/\sigma_N < 3$) is $\sim$1.3, slightly larger than the expected value of
one.  This may indicate either that among the stars without significant null excess there are still stars with tentative excesses, or that we
slightly underestimate our measurement uncertainties.  While the former can generally be expected, the latter is supported by the symmetrical
distribution of non-detections around $N_{\text{as}}/\sigma_N = 0$.  The distribution of the measurement uncertainties is well behaved with a
sharp peak at low uncertainty and a tail toward higher uncertainties for stars observed under less suitable conditions or for which a smaller
amount of data was obtained than for others.  As expected from background photon and detector read noise, the median null uncertainty
increases with aperture size.  The larger scatter for the conservative aperture can be explained by the fact that this aperture is optimized
for each star and thus its size (and with it the photon and read noise of the measurement) changes from target to target.  Based on these
arguments, we define a significant excess detection as a star for which we measure $N_{\text{as}}/\sigma_N > 4$ in at least one aperture.

In principle, any of our detections could be caused by the presence of an unknown binary companion instead of a dust disk.  However, most of
our targets have been observed at a range of parallactic angles.  A binary companion would rotate across the transmission pattern in and out
of the transmissive fringes with parallactic angle rotation.  This will typically result in a variation between full null excess (companion
on transmissive fringe) and zero null excess (companion on dark fringe), while limited field rotation may result in any scenario in between
these extrema depending on the exact configuration of a binary system.  Although the limited excess significance for most of our detections
prevents a definitive conclusion, binarity is an unlikely scenario.  Many of our targets have also been observed with high contrast imaging
observations searching for giant planets and no detections of binary companions have been reported \citep[e.g.,][]{stone2018, mawet2019}.

We have detected significant excesses around 10 stars out of the total of 38 stars observed.  In fact, all these stars show excesses
$N_{\text{as}}/\sigma_N > 5$ and/or have been detected combining consistent data from at least two independent observations (i.e., in at
least two different nights).  We thus consider all these detections robust.

\subsection{Null-to-zodi conversion}
\label{sect_null-to-zodi}

A detailed description of the modeling strategy for the HOSTS data has been presented by \citet{kennedy2015a} and updated by
\citet{ertel2018a}.  In Appendix~\ref{app_modeling} we provide a cookbook on how to compare a disk model to our null measurements for general
model fitting.

For the conversion from astrophysical null measurements to dust levels (zodis), we used the model presented by \citet{kennedy2015a}.  It
describes a radial dust surface density distribution analogous to the Solar system's zodiacal dust \citep{kelsall1998}, scaled in size with
the square root of the host star's luminosity.  We scale the dust surface density (vertical geometrical optical depth) of this model to
7.12$\times10^{\text{-8}}$ at the EEID, equal to the surface density of the zodiacal dust at 1\,au from the Sun \citep{kelsall1998}.  This
defines the unit of 1\,zodi which we use to quantify the HZ dust levels around our target stars.

Note that the unit of 1\,zodi is a unit of vertical geometrical optical depth (surface density) of the dust in a star's HZ.  It thus does
not depend on the observing wavelength or emission mechanism.  We emphasize here that there are limitations to this approach related to the
simplifications of the model and the likely variety of planetary system and dust architectures around our target stars.  In particular, if the
spectral shape of the dust emission is different from the Solar system's, applying our this method of measuring a star's zodi
level to observations at a different wavelength would yield a different zodi measurement result.  For detailed discussions of the
shortcomings of our approach and how they are at least in part mitigated by the optimized design of the LBTI we refer to \citet{kennedy2015a}
and \citet{ertel2018a}.

The usually unknown orientation of the potential dust disk (inclination and position angle) were randomized and the response of the LBTI to
all possible orientations was used to compute a most likely null-to-zodi conversion factor (the astrophysical null $N_{\text{as,1}}$ expected
from a 1\,zodi disk, Table~\ref{tab_measurements}).  As pointed out by \citet{ertel2018a}, in practice the uncertainty from the disk
orientation is negligible compared to the null measurement uncertainty due to the range of hour angles over which each target has been
observed.  Correction factors for the finite aperture size were computed from the same model by convolving the model image of the
transmitted dust emission with the single aperture point spread function of the observations and dividing the total predicted null excess
from the model by the null excess predicted in a given aperture.  For detected excesses we converted the astrophysical null measurement from
the aperture that yields the most significant detection (Table~\ref{tab_measurements}) to a zodi level.  For non-detections we used the
measurement based on the noise optimized aperture assuming a dust distribution analogous to the Solar system's zodiacal dust.  All our
detections agree with this assumption within the measurement uncertainties.  We find a median 1\,$\sigma$ sensitivity of 23\,zodis for early
type stars and 48\,zodis for Sun-like stars.

\setcounter{table}{2}
\begin{deluxetable}{cccc}
\tablecaption{Subsamples, LBTI excess detections and rates vs.\ auxiliary data.\label{tab_rates}}
\tablecolumns{4}
\tabletypesize{\normalsize}
\tablehead{
 ~~~~~~~~~~~~ & ~~Early type~ & ~~Sun-like~~ & ~~~~~All~~~~~
}
\startdata
\hline
 All                       & 6 of 15            & 4 of 23            & 10 of 38           \\[-4pt]
 stars                     & $40^{+13}_{-11}\%$ & $17^{+10}_{-5}\%$  & $26^{+8}_{-6}\%$   \\[2pt]
\hline
 Cold                      & 5 of 6             & 2 of 3             & 7 of 9             \\[-4pt]
 dust                      & $83^{+6}_{-23}\%$  & $67^{+15}_{-28}\%$ & $78^{+8}_{-18}\%$  \\[4pt]
 No cold                   & 1 of 9             & 2 of 19            & 3 of 28            \\[-4pt]
 dust                      & $11^{+18}_{-4}\%$  & $11^{+11}_{-4}\%$  & $11^{+9}_{-3}\%$   \\[4pt]
\hline
 Hot                       & 3 of 6             & 1 of 2             & 4 of 8             \\[-4pt]
 excess                    & $50^{+18}_{-18}\%$ & $50^{+25}_{-25}\%$ & $50^{+16}_{-16}\%$ \\[4pt]
 No hot                    & 3 of 7             & 1 of 13            & 4 of 20            \\[-4pt]
 excess                    & $43^{+18}_{-15}\%$ & $8^{+14}_{-3}\%$   & $20^{+12}_{-6}\%$  \\[4pt]
\hline
\multirow{2}{*}{Young$^b$} & 5 of 8             & 3 of 12            & ... \\[-4pt]
                           & $63^{+13}_{-18}\%$ & $25^{+15}_{-8}\%$  & ... \\[4pt]
\multirow{2}{*}{Old$^b$}   & 1 of 8             & 1 of 12            & ... \\[-4pt]
                           & $13^{+20}_{-4}\%$  & $8^{+15}_{-3}\%$  & ... \\[4pt]
 Young                     & 4 of 7             & ... & ... \\[-4pt]
 w/o $\zeta$\,Lep          & $57^{+15}_{-18}\%$ & ... & ... \\[4pt]
 Old                       & 0 of 7             & ... & ... \\[-4pt]
 w/o $\eta$\,Crv           & $0^{+21}_{-0}\%$   & ... & ... \\[2pt]
\hline
\enddata
\tablecomments{The presence or absence of cold dust and hot excess for our target stars is indicated in Table~\ref{tab_stars}.\\
$^b$~Stars younger or older than the median age of their respective spectral type bin.  The star with the median age in each subsample (a
non-detection in each case) was included in both the young and old group, which is why the sum of young and old stars is one larger than the
total number of stars.}
\end{deluxetable}

\section{Discussion}
\label{sect_disc}

In this section we interpret our results.  We first discuss the detection rates and their correlations with other system parameters and
hypothesize about the sources of the correlations (Sect.~\ref{sect_disc_corr}).  We then briefly discuss the potential for further
observations and detailed analyses of our strong detections to better understand these individual systems (Sect.~\ref{sect_disc_followup}). 
A statistical analysis to derive the typical zodi level around the Sun-like stars and a discussion of the implications for future exo-Earth
imaging, including the merit of more observations with an improved sensitivity that can realistically be achieved by moderate instrument
upgrades to the LBTI is presented in Sect.~\ref{sect_disc_sample}.

\subsection{Detection statistics and correlation with other system parameters}
\label{sect_disc_corr}

We detect significant excesses around 10 stars out of the total of 38 stars observed.  These detections include $\beta$\,Leo (Defr\`ere
et al., in prep.) and $\eta$\,Crv \citep{defrere2015}.  We previously excluded those two targets from the statistical analysis
of an early subset of HOSTS observations in \citet{ertel2018a}, because the data on them were taken during commissioning time, not as part of
the unbiased HOSTS survey.  Here we assume that toward the end of the HOSTS survey, as the number of available targets that had not yet been
observed decreased, both stars would have been observed by the unbiased survey if they had not been observed as commissioning targets.  They
are thus now considered part of the unbiased survey.  We will see that our detection statistics are consistent with these from
\citet{ertel2018a}, so no significant bias is introduced from including or excluding these two stars.

The basic detection statistics for different subsamples of targets are summarized in Table~\ref{tab_rates}.  Our sample size is limited and
any statistical analysis is affected by large statistical uncertainties and small number statistics.  We illustrate this by displaying
binomial uncertainties with our detection rates.  In addition, while the accuracy of our null measurements is independent of stellar spectral
type, it is not the same for every star due to differences in data quality and quantity of individual targets.  Moreover, our sensitivity to
HZ dust is limited (as quantified by the sensitivity to dust in units of zodi) and decreases from earlier to later stellar spectral
types \citep{kennedy2015a}.  As a consequence, our detection rates cannot readily be converted into occurrence rates.  Caution must be
exercised when interpreting our detection rates and any theoretical work predicting occurrence rates of exozodiacal dust needs to be compared
to our observations for the individual stars directly rather than the detection rates.  Such theoretical work is beyond the scope of the
present paper.  We thus limit ourselves in the following to a qualitative discussion of our detection rates and compare them to a range of
other properties of the systems to search for correlations.

\subsubsection{No correlation with stellar spectral type}

\begin{figure}
 \centering
 \includegraphics[angle=0,width=\linewidth]{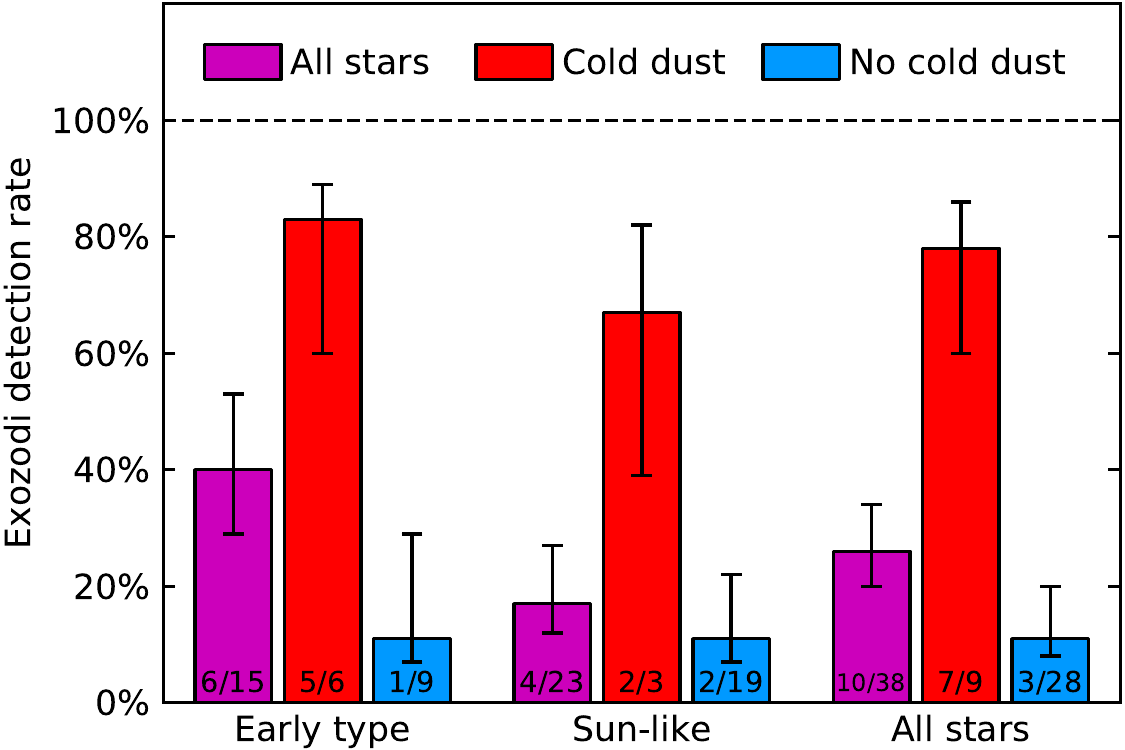}
 \caption{Histogram of HZ dust detection rates with respect to stellar spectral type and the presence of a detected, cold debris disk.  The
   correlation between detection rate and spectral type is likely related to a sensitivity bias (Sect.~\ref{sect_disc_corr}).  The
   correlation between detection rate and presence of cold dust is likely of astrophysical origin and found to be significant for early type
   stars, while no conclusion can be drawn for Sun-like stars.  The number of stars and detections in each subsample is indicated at the
   bottom of the corresponding bar.}
 \label{fig_hist_dusty_clean}
\end{figure}

Our detection statistics with respect to stellar spectral type are shown in Fig.~\ref{fig_hist_dusty_clean}.  We find a higher over-all
detection rate for early type stars than for Sun-like stars of $40^{+13}_{-11}\%$ and $17^{+10}_{-5}\%$, respectively.  Closer investigation,
however, shows that this trend simply illustrates the dominant bias in our survey:  It can be explained entirely by the spectral type
dependence of the LBTI's sensitivity.  If we observed the zodi levels $z$ measured around our early type stars with the sensitivity
$\sigma_z$ to HZ dust of our Sun-like stars, we would expect a detection rate of 18\% (4 of 23 stars), identical to our observed detection
rate for Sun-like stars.  We thus see no evidence in our data of a correlation of the occurrence rate or amount of HZ dust with stellar
spectral type.

\subsubsection{No correlation with stellar age}

\begin{figure}
 \centering
 \includegraphics[angle=0,width=\linewidth]{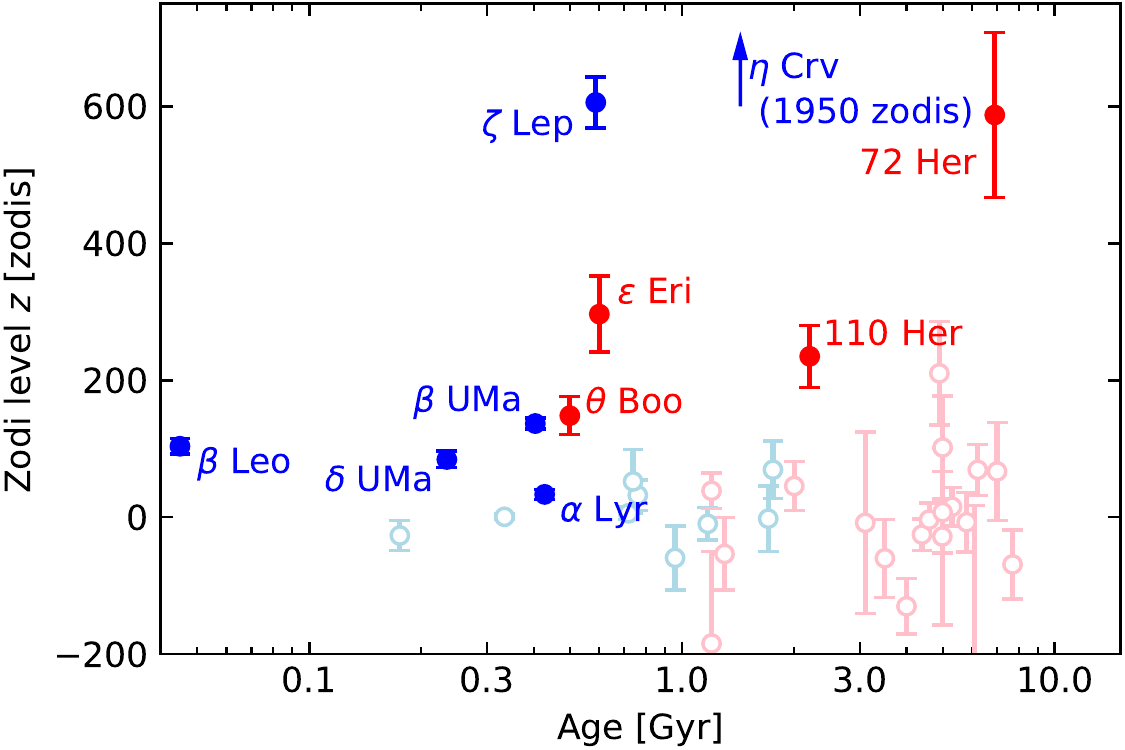}
 \caption{Age distribution of our target stars and the corresponding zodi measurements.  \emph{Blue} dots are for early type stars,
   \emph{red} ones for Sun-like stars.  \emph{Filled} symbols are for LBTI detections, \emph{open, faint} circles are for non-detections.}
 \label{fig_ages}
\end{figure}

Fig.~\ref{fig_ages} shows the zodi levels of our targets vs.\ stellar age.  Ages for the sun-like stars were taken from the compilations by
\citet{gaspar2013} and \citet{sierchio2014}.  Four of the stars had relatively weak determinations in those works:  13\,UMa, 40\,Leo,
61\,Cyg\,A, and $\iota$\,Psc.  We checked the ages for them against all relevant work since those papers were published and confirmed them
for three, but a significantly different age of 7.0\,Gyr has been found for 61\,Cyg\,A by astroseismology \citep{metcalfe2015} and adopted
here.  Ages for the early-type stars are based on the modes in the 1D fits by \citet{david2015}.  Their determinations were from isochrones,
a technique that loses resolution for young stars near the zero age main sequence. We therefore checked such stars against other sources,
finding general consistency except for $\beta$\,Leo, which \citet{zuckerman2019} finds to be a member of the Argus moving group with an age
of 40 to 50\,Myr.  Except for this latter case, where we adopted the moving group age, we used the ages from \citet{david2015}, so our
comparisons would be on a consistent scale.

\begin{figure}
 \centering
 \includegraphics[angle=0,width=\linewidth]{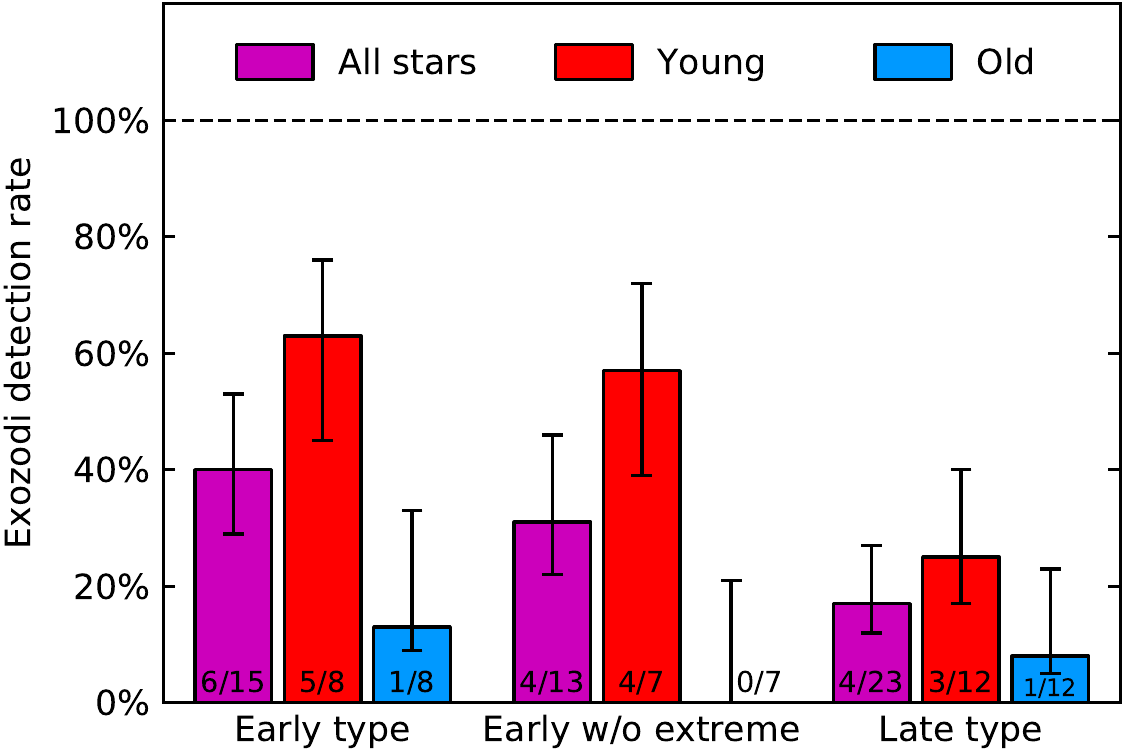}
 \caption{Detection rates with respect to the stellar age measured with respect to the median age of the sample a star belongs to.  The
   visible trends can be attributed to a sensitivity bias (Sect.~\ref{sect_disc_corr}).  The number of stars and detections in each
   subsample is indicated at the bottom of the corresponding bar.  The star with the median age in each subsample (a non-detection in each
   case) was included in both the young and old group, which is why the sum of young and old stars is one larger than the total number of
   stars.}
 \label{fig_hist_ages}
\end{figure}

We see in Fig.~\ref{fig_ages} that the stars with detected LBTI excesses tend to be on the younger side of both the early type and the
Sun-like samples with a few exceptions.  When separating the two samples into stars younger and older than the median age of the respective
sample (718\,Myr for the early type stars, 4.6\,Gyr for the Sun-like stars), we find a higher detection rate for younger stars than for
older ones (Fig.~\ref{fig_hist_ages}).  This correlation becomes even clearer if we exclude the two potentially extreme cases of $\eta$\,Crv
and $\zeta$\,Lep.  This is, however, likely a result of the same bias with stellar spectral type discussed in the previous section.  Stars of
earlier spectral type have shorter life times than stars of later spectral type.  Thus, the early type and Sun-like stars older than the
median age of their respective spectral type samples are on average of later spectral types than stars younger than the median age of their
respective spectral type sample.  As we are less sensitive for stars of later spectral type, we are on average less sensitive for stars in the
older age bin than those in the younger age bin for each spectral type sample.  If we observed the excesses measured around the younger stars
in each spectral type sample with the sensitivities of the older stars in the same sample, we would expect detection rates that are marginally
higher than -- but entirely consistent with -- those found for the older stars in each sample.  Our small sample sizes prevent us, however,
from seeing weak trends.

What is potentially more enlightening is the fact that the strongest excesses are not detected around the youngest stars.  Among the early
type stars the extreme excess around $\eta$\,Crv stands out due to the Gyr age of the star.  The strong detection around the intermediate age
early type star $\zeta$\,Lep may or may not be another such case, but this large excess may also be caused by the proximity of the `cooler'
dust in this system to the HZ (see Sect.~\ref{sect_trend_cold_dust}).  Among the Sun-like stars the cases of 72\,Her and 110\,Her at ages of
several Gyr show that strong excesses may be present at any stellar age.  Our detections at ages well beyond the $\sim$500\,Myr lifetime of
24\,$\mu$m excesses \citep{gaspar2014} suggest that the HZ dust in these systems is not linked to (decayed) asteroid belts, but may either
arise from recent stochastic events or be linked to outer cold disks with longer life times \citep{sierchio2014}.

\subsubsection{Strong correlation with the presence of cold dust}
\label{sect_trend_cold_dust}

A strong correlation is visible in Fig.~\ref{fig_hist_dusty_clean} between the zodi detection rate and the presence of a known outer debris
disk, detected photometrically through the far-infrared excess it produces around its host star.  For the majority of our target stars with a
known cold debris disk (seven of nine, $78^{+8}_{-18}$\%) we have also detected HZ dust, while only three of 28 stars without cold dust have
detected HZ dust.  We use the $p$-value from Fisher's exact test to evaluate if this correlation is significant.  This is justified here
despite the non-uniform sensitivity across our sample, because the sensitivity does not directly depend on the presence or absence of cold
dust, so that this property does not introduce any bias.  The correlation is strong for early type stars ($p=0.01$), while for Sun-like stars
the small number of three known debris disks in our sample prohibits a definite conclusion ($p=0.07$).  Observing stars with debris disks was
not a priority of the HOSTS survey as such stars are unlikely to be first choice targets for future exo-Earth imaging missions.  Thus, the
HOSTS samples were designed to be unbiased with respect to the presence of a cold debris disk.  Since detectable debris disks are less common
around Sun-like stars than around Early-type stars \citep{rieke2005, montesinos2016}, few stars in our Sun-like sample host such disks.

The correlation between our HZ dust detections and cold dust suggests that the origin of bright HZ dust is somehow connected to the presence
of dust or minor bodies further away from the star, e.g., through inward transport of dust due to PR drag \citep{wyatt2005} or through dust
delivery by comets \citep{nesvorny2010, faramaz2017, sezestre2019}.  It is, however, noteworthy that there are several detections of HZ dust
in systems that do not have a detected cold debris disk despite sensitive searches.  This may suggest an alternative origin of the dust in
these systems or that even cold debris disks that are too faint to be detected by current methods may still be a significant source of
exozodiacal dust \citep{bonsor2012, bonsor2014}.  It is important to point out here the two uncertain cases of $\theta$\,Boo and
110\,Her.  Both systems have tentative detections of cold dust.  We consider the far-infrared excess around 110\,Her significant as it has
been detected independently with Spitzer at 70\,$\mu$m and Herschel at 70\,$\mu$m and 100\,$\mu$m, albeit with marginal significance,
but consider $\theta$\,Boo a non-detection with only a 2.5\,$\sigma$ excess found by Herschel.  Moving either of these two stars to the other
category (cold excess vs.\ no cold excess or vice-versa) would not change our conclusions, but this illustrates that our analysis is limited
not only by our own data but also the sensitivity of available debris disk surveys.  The Herschel non-detection for 72\,Her is not very
constraining with the strongest upper limit at only 40\% of the stellar photosphere at 100\,$\mu$m \citep{eiroa2013}.

\begin{figure}
 \centering
 \includegraphics[angle=0,width=\linewidth]{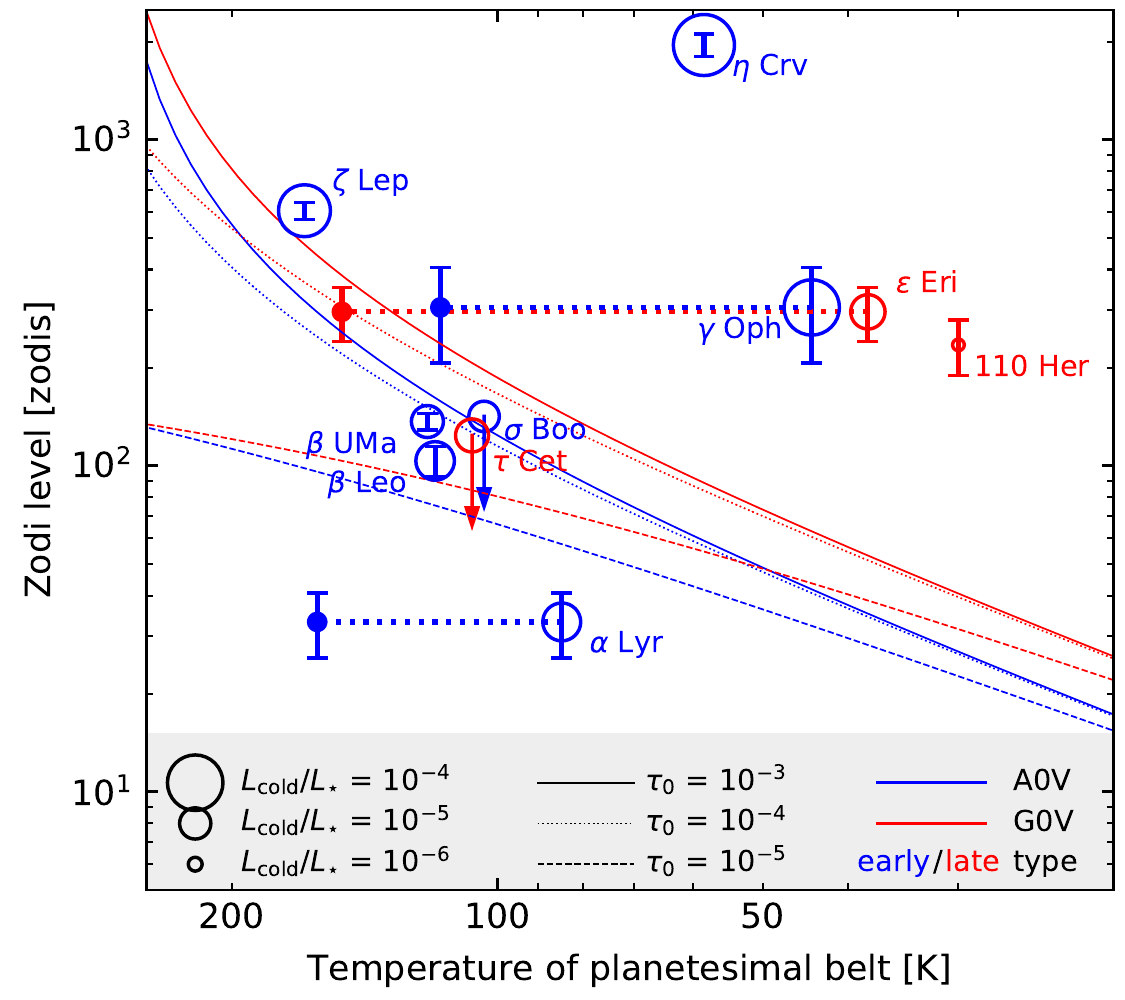}
 \caption{Zodi levels vs.\ temperature and luminosity of outer debris disks (where detected).  \emph{Blue} symbols are for early type stars,
   \emph{red} ones for Sun-like stars.  Downward arrows indicate upper limits on zodi levels.}
 \label{fig_temp_lum}
\end{figure}

Fig.~\ref{fig_temp_lum} shows our measured zodi levels with respect to the temperature and fractional luminosity of the cold, outer debris
disk (measured by a single modified blackbody fit to the spectral energy distribution of the far-infrared to millimeter excess measurements
from the literature) for systems for which such a disk has been detected (7 out of our 10 detections and two of our non-detections).  We also
add $\gamma$\,Oph from \citet{mennesson2014}.  For stars with well known warm belts inside the cold, outer belts we include a point at the
temperature of the warm belt, too.  In addition, we plot model predictions of the amount of dust delivered to the HZ from outer belts at
various temperatures and vertical optical depths under the influence of PR drag and collisions.  The vertical optical depth of the dust in
the HZ is computed following the equation (adapted from \citealt{wyatt2005})
\begin{equation}
  \tau_{\rm{HZ}} = \frac{\tau_0}{1 + 4\times10^4 \times \tau_0 \frac{278\,\rm{K}}{T_0} \left(\frac{L_\star^{0.25}}{M_\star^{0.5}}\right)
     \left(1-\frac{T_0}{278\,\rm{K}}\right)}
\end{equation}
where $T_0$ and $\tau_0$ are the temperature and vertical optical depth of the outer belt, respectively, and the stellar luminosity $L_\star$
and mass $M_\star$ are measured in Solar units L$_\odot$ and M$_\odot$.  The zodi level is then $z = \tau_{\rm{HZ}}\,/\,7.12\times10^{-8}$
following our definition in Sect.~\ref{sect_null-to-zodi}.  This plot is analogous to Fig.~10 in \citet{mennesson2014}, but plotted in zodi
level instead of null depth and showing lines for various $\tau_0$ (the lines in \citet{mennesson2014} are shown for $\tau_0=10^{-4}$).

The predictions of the radial surface density distribution from this model are not identical to our Solar zodi model used to convert our null
measurements to zodis, which means that the two are not fully compatible (Sect.~\ref{sect_null-to-zodi}).  However, both models have a fairly
flat radial distribution throughout the HZ and the design of the LBTI partly mitigates the impact of this discrepancy.  The model has also
been noted to under predict the effect of collisions and thus over predict the $N$~band flux of the disk \citep{kennedy2015b}.

Despite those caveats, it is noteworthy that the model predicts most of our zodi measurements reasonably well.  Because the model is unlikely
to under predict the HZ dust level, perhaps the strongest conclusions possible are that the HZ dust of $\eta$\,Crv cannot be explained by this
model and that the HZ dust in the $\gamma$\,Oph and $\epsilon$\,Eri systems is more likely to originate from the warmer belts than the outer,
cold belt if produced by PR drag (but stellar wind drag may affect the latter conclusion for low luminosity stars,
\citealt{reidemeister2011}).  Another potential outlier is 110\,Her, but the detection of cold dust is very weak and the constraints on
warmer dust in the system are relatively poor \citep{eiroa2013}.  Thus, the system could be similar to $\epsilon$\,Eri with an Asteroid belt
analog that could be responsible for the large amounts of HZ dust.  Furthermore, by comparison with the other detections around early
A-type stars ($\beta$\,Leo, $\beta$\,UMa, and $\zeta$\,Lep) the zodi level of $\alpha$\,Lyr appears very low which may suggest clearing by
planets \citep{bonsor2018} in or outside the HZ if the dust in all of these systems is delivered by PR drag.  Finally, our observations are
consistent with the model predictions of closer outer belts producing higher zodi levels and little spectral type dependence of this effect,
but our small number statistics do not allow for strong conclusions.

\subsubsection{No connection with hot dust}

\begin{figure}
 \centering
 \includegraphics[angle=0,width=\linewidth]{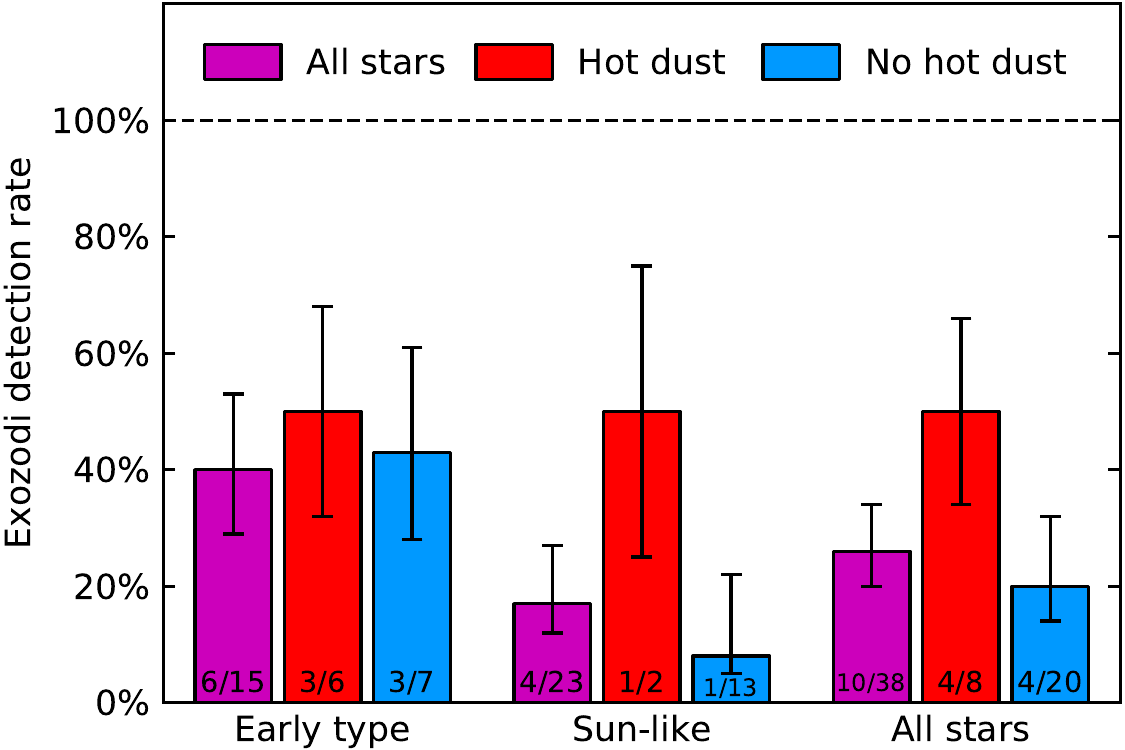}
 \caption{Histogram of HZ dust detection rates with respect to the presence of a detected near-infrared excess.  The number of stars and
   detections in each subsample is indicated at the bottom of the corresponding bar.}
 \label{fig_hist_hot_dust}
\end{figure}

We see no correlation between the presence of hot and HZ dust around early type stars (Fig.~\ref{fig_hist_hot_dust}).  The
difference for Sun-like stars is not significant either ($p=0.26$).

\subsubsection{A consistent picture from the present and absent trends}

It is possible that we see the signs of different dust origins:  For the early type stars, for which we are the most sensitive, we may
be able to detect the results of a delivery of material from an outer debris disk in some sort of continuous process.  PR drag or a steady
flow of comets from the outer system to the inner regions are potential mechanisms for this delivery and both are likely at play to some
degree depending on the architecture of each system.  This would correlate with the presence of a cold disk and the delivered amount
of dust would potentially decrease over time:  \citet{wyatt2005} has shown that the HZ dust level for the PR drag scenario depends only
weakly on the mass of the outer disk, and thus the effects of decreasing debris disk masses with age may be small (but measurable for low
outer disk optical depths and warm outer disks, Fig~\ref{fig_temp_lum}).  \citet{faramaz2017} have
shown that comet delivery depends on the number of large bodies on suitable orbits in the outer system which decreases over time due to their
removal by both debris disk evolution and their ongoing delivery to the HZ.  However, they have also found that HZ dust disks may be
sustained over Gyr time scale.  On the other hand, there are potentially extreme systems such as $\eta$\,Crv.  These systems may be produced
by sporadic, catastrophic events and as illustrated by the Gyr age of $\eta$\,Crv, these events may occur at least at this age.  For Sun-like
stars we may typically be only sensitive enough to detect such extreme systems.  Such events would also not necessarily originate from and
correlate with the presence of a detectable debris disk, explaining our detections in systems without cold dust.  Alternatively,
\citet{bonsor2012} and \citet{marino2018} have suggested specific planetary system architectures that may support a high influx of comets.

These hypotheses may be tested observationally.  Improving the sensitivity of the LBTI by a factor of two to three is realistic with
moderate instrument upgrades (Sect.~\ref{sect_sensitivity}).  This should allow for the detection of more of the supposedly continuously
supplied HZ dust systems around early type stars and for testing the expected correlations with outer disk mass and temperature.  It may also
allow for the detection of such systems around Sun-like stars.  The detailed study of the detected systems with the LBTI
\citep{ertel2018spie} and current and future instruments on the Very Large Telescope Interferometer \citep{ertel2018b, kirchschlager2018,
defrere2018} may also allow us to determine the origin of the dust in these individual cases and the connection between the HZ dust and
hotter dust even closer to the stars, thus helping to understand the origins and dynamics of the various dust species in the HZs of the stars
and closer in.

\subsection{Potential for the detailed study of specific targets}
\label{sect_disc_followup}

In addition to the statistical constraints derived from the HOSTS observations, the data also provide important constraints on specific
systems for which exozodiacal dust has been detected or for which strong and interesting upper limits have been found.  $\beta$\,UMa
and $\beta$\,Leo are examples of relatively strong HZ dust detections in systems with known cold dust.  In contrast, $\alpha$\,Lyr has a
rather low zodi level despite a massive cold disk, which may be explained by the large size of the outer disk (e.g., given the possible
correlation in Fig.~\ref{fig_temp_lum}) or the presence of a giant planet preventing dust from migrating inward in case of the PR drag
scenario \citep{bonsor2018}.  $\epsilon$\,Eri is a nearby, interesting late type star for exo-Earth imaging, but has a very high HZ dust
level which will complicate planet detection.  On the other hand, it seems to be the ideal target for studying planet-disk interaction in the
HZ.  Furthermore, it might be the only Sun-like star in our sample with detected, continuously supplied HZ dust, making it a prototype for
studying the relative importance of PR/SW drag and comet delivery around Sun like stars.  The warm dust in the 110\,Her system seems to be
concentrated relatively far from the star\footnote{While all our detections are consistent with a moderately increasing excess with aperture
size, as expected from a dust distribution analog to the Solar system's zodiacal dust, 110\,Her's excess is strongly increasing with
photometric aperture size and is the only case with no significant detection in the 8\,pix aperture and 13\,pix apertures, but a detection in
the conservative aperture.  Given the large uncertainties, it is however not clear if this is significant, so that this needs to be
investigated further by a deeper analysis of the available data and new observations.}, while the large amount of dust around $\eta$\,Crv is
located very close in \citep{defrere2015}.  If a catastrophic event has produced the dust in both systems, this will hint at the separation
at which this event occurred.  In the PR drag scenario, the dust location around 110\,Her could hint at the presence of a massive planet just
inside that separation preventing the dust from migrating further in \citep{bonsor2018}.  Several systems have high HZ dust levels despite
the lack of detected cold dust, which may complicate the target selection for exo-Earth imaging and needs to be understood. 
Furthermore, in addition to the detected HZ dust, several systems also have hot dust such as $\alpha$\,Lyr \citep{absil2006} and
$\beta$\,Leo \citep{absil2013}, while others such as $\epsilon$\,Eri do not.  Such systems may allow us to further study the connection
between the warm and hot dust and to place additional constraints on the origin of both and the architectures of the planetary systems around
those stars.

Our detections can be studied in detail to understand their properties and the diversity of their architectures, and to support our
interpretation of the correlations discussed in the previous section between the HZ dust level and other properties of a system.  Such
studies also improve our understanding of the formation and evolution of the HZ dust and thus increase the ability of models to predict the
level of HZ dust in systems that could not be observed by the HOSTS survey.  This will critically assist in the target selection for future
exo-Earth imaging missions.

We have performed the first such studies on the existing data (\citealt{defrere2015}, Defr\`ere et al. in prep.) and the HOSTS science team
is currently analyzing the most relevant, remaining detections.  Follow-up observations with the LBTI at a wide range of position angles and
different wavelengths are critical, however, to derive strong constraints on the architectures of the detected dust disks
\citep{ertel2018spie}.  Detailed modeling of these data together with available literature data \citep[e.g.,][]{ertel2011, ertel2012a,
lebreton2013, ertel2014a, lebreton2016} can be used to create a comprehensive picture of each system and to predict its appearance at other
wavelengths  (e.g., the HZ dust brightness in scattered light).  Improving the sensitivity of the LBTI will provide higher quality data for
even stronger constraints.  Furthermore, the wider community has already taken up the first HOSTS publications for further analysis.
\citet{bonsor2018} have developed a model to predict or rule out the presence of giant planets in a system based on the mass and location of
an outer belt and the level of HZ dust from our observations.  More analyses of our detections will help calibrating this model to produce
accurate constraints.

\begin{figure*}
 \centering
 \includegraphics[angle=0,width=\linewidth]{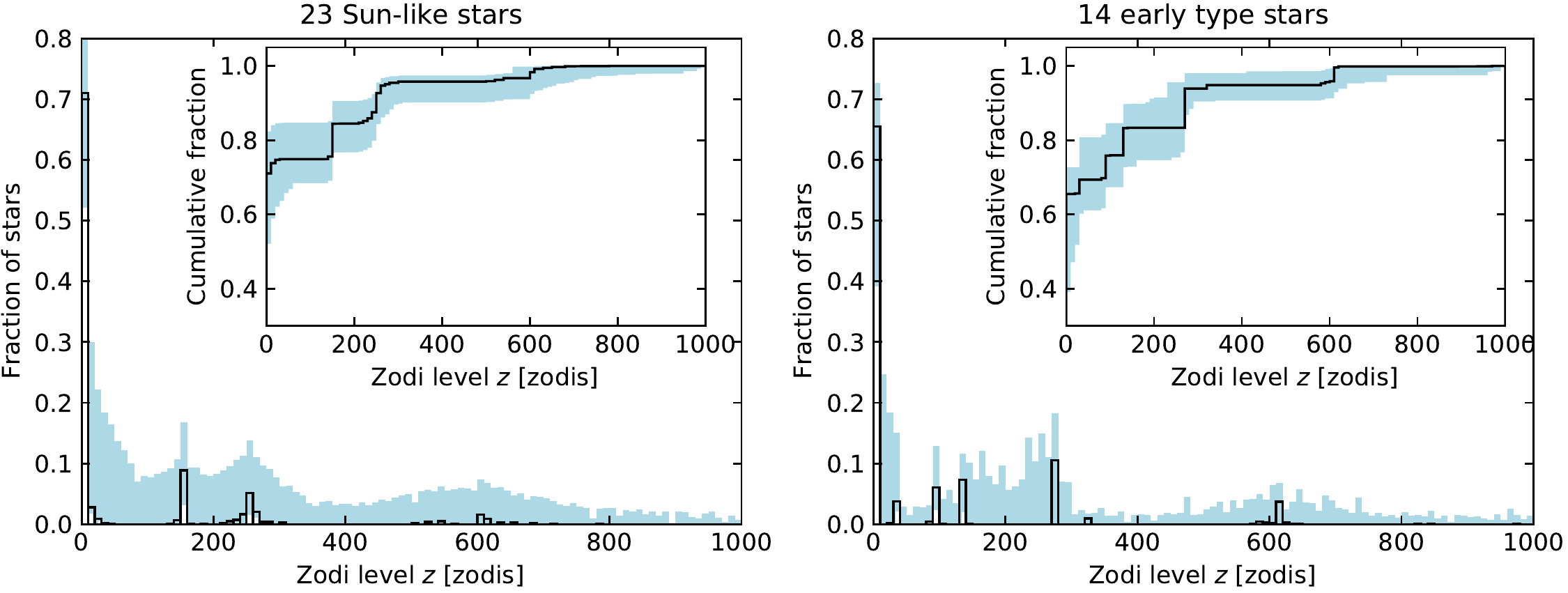}
 \caption{Best-fit free form luminosity function fit to the HOSTS data for Sun-like stars (\emph{left}) and early-type stars (\emph{right}). 
   Both the fraction of stars in each zodi level bin and the cumulative distribution are shown.  The black line in each plot shows the best
   fit distribution, while the blue bars show the 1\,$\sigma$ range for each zodi bin.  The original 1 zodi bins used for our analysis are
   increased to larger bins of 10\,zodis for better visualization and reduction of statistical noise in the images.}
 \label{fig_dist}
\end{figure*}

\section{Sample constraints on habitable zone dust and implications for exo-Earth imaging}
\label{sect_disc_sample}

The primary objective of the LBTI has been to inform the design of a future exo-Earth imaging space telescope mission.  LBTI's mission
success criteria describe a desire for `a high confidence prediction of the likely incidence of exozodi dust levels above those considered
prohibitive' for such a mission.  In this section we perform a statistical analysis to answer this question, discuss the implications of our
results for future exo-Earth imaging and characterization missions, and outline a path for further improvements.

\subsection{Sample constraints on habitable zone dust}
\label{sect_free_form}
In our previous analysis of an early subset of HOSTS observations \citep{ertel2018a}, we assumed a log-normal distribution of the fraction of
stars at a given zodi level (luminosity function) and fitted it to our zodi measurements for different subsamples of stars to determine the
median zodi levels of these samples and their uncertainties.  We found that:  (1) a lognormal luminosity function appears inadequate to
reproduce the observed distribution of excesses well, instead a bimodal luminosity function in which most stars have low zodi levels and a
few `outliers' have relatively high levels is more likely, and  (2) within our statistical uncertainties, there is no significant difference
(using Fisher's exact test) between Sun-like stars with and without cold dust as is seen for early-type stars.  The former is further
supported by our complete survey data, while the latter remains valid.  While we see a clear correlation between the detections of cold and
HZ dust in our over-all sample, the statistics are not good enough to confirm the tentative correlation for Sun-like stars.  In particular,
the fact that we find LBTI excesses for stars without known debris disks shows that Sun-like stars without far-infrared excesses do not
constitute a clean sample of stars with low HZ dust levels.  Thus, we do not distinguish between stars with and without detected cold excess
for our luminosity function analysis.  Because the lognormal distribution is not a good fit to our data and there is reason to believe that
a single mechanism inadequately describes the dust production, we use the `free-form' iterative maximum likelihood algorithm described by
\citet{mennesson2014} instead.

For the free-form method, the explored zodi levels for the two spectral type samples, respectively, are binned and the unknown luminosity
function is parameterized through the fraction of stars that have a zodi level in each of the bins.  For our analysis we selected bins of
equal width of 1\,zodi ranging from 0\,zodis to 2000\,zodis, an upper boundary consistent with the LBTI measurements of all stars other than
$\eta$\,Crv.  We excluded the latter star as a clear and extreme outlier to limit the computational effort of our analysis.  The fraction of
stars in each bin is then adjusted iteratively to maximize the likelihood of observing the data (\citealt{mennesson2014}).  The median zodi
level $m$ was used to characterize the distribution.  To determine the uncertainty of the derived distribution, we disturbed this `nominal'
distribution, creating 10$^\mathsf{5}$ new distributions with small deviations from the nominal one.  The likelihood of observing the data
was computed for each of these distributions, and the profile likelihood theorem was then used to derive 1$\sigma$ confidence intervals on
$m$ from its distribution among them.  We derive a median zodi level of $m = 3^{+6}_{-3}$\,zodis (95\% upper limit: 27\,zodis) for Sun-like
stars and $m = 2^{+28}_{-2}$\,zodis (95\% upper limit: 53\,zodis) for early-type stars based on the zodi values derived following
Sect.~\ref{sect_null-to-zodi}.  The uncertainties on the fraction of stars in each bin of the luminosity function were derived as the range
of values encountered for each bin among the distributions that fall within 1\,$\sigma$ and 95\% probability of the best-fit distribution.  
The higher upper uncertainties on the median for the early-type stars despite the smaller uncertainties of the individual zodi measurements
can be explained by the smaller number of stars and the fact that a larger fraction has significant detections above the best-fit median zodi
level.

As an experiment, we re-computed the statistics for our sample of Sun-like stars, but adding the Sun itself with a zodi level defined to be
1\,zodi.  This did not significantly change our results, which is unsurprising as our results are entirely consistent with one star out of
24 Sun-like star having a zodi level of exactly 1\,zodi.

Histograms of the best-fit free form distributions and their $1\,\sigma$ ranges are shown in Fig.~\ref{fig_dist}.  It is remarkable how
similar the distributions are for early-type and Sun-like stars despite the higher detection rate and higher fraction of early type stars
with cold dust compared to Sun-like stars.  This further reinforces our earlier conclusion that there is no significant difference in our
data between the two spectral type samples that cannot be explained by the different sensitivity to zodi levels of our observations.

Our results for Sun-like stars are recommended for the yield calculation of future exo-Earth imaging missions and have been adopted by the
Habitable Exoplanet Observatory (HabEx; \citealt{gaudi2018}) and Large UV-Optical InfraRed Surveyor (LUVOIR; \citealt{luvoir2018}) mission
study teams as well as for the Large Interferometer For Exoplanets (LIFE) concept study \citep{quanz2018}. Histograms of the best-fit free
form distribution and its $1\,\sigma$ range are shown in Fig.~\ref{fig_dist}.

\subsection{Implications for exo-Earth imaging}
\label{sect_imp_exoearths}

Direct imaging blends the light of an exoplanet with the light scattered by any surrounding exozodiacal dust.  The amount of exozodi
contamination is proportional to the sky area of the photometric aperture being used, which in turn depends on the telescope's
diffraction-limited beam size.  Large telescopes have more compact PSFs and mix less exozodi signal in with the exoplanet signal, while
smaller telescopes have larger PSFs that result in more blending of unwanted exozodi signal with planet light. For any given telescope, the
exozodi contamination in exoplanet images is worse at longer wavelengths due to the larger diffraction-limited  beam size, and worse for more
distant targets where the exoplanet signal is fainter but the exozodi surface brightness is unchanged.

The primary objective of LBTI has been to inform the design of a future exo-Earth imaging space telescope mission.  LBTI's mission success
criteria describe a desire for `a high confidence prediction of the likely incidence of exozodi dust levels above those considered
prohibitive' for such a mission.  The science and instrument requirements defined at the 2015 start of the survey were derived from this
consideration, given the best available knowledge at that time of the impact of exozodiacal dust on such missions.  Since then, three mission
concept studies have been developed that include exo-Earth direct detection as a major objective:  The WFIRST Starshade Rendezvous Probe
\citep{seager2019}, the Habitable Exoplanet Observatory \citep[HabEx;][]{mennesson2019c} and the Large UV-Optical InfraRed Surveyor
\citep[LUVOIR;][]{roberge2019}.  At the same time there has been ongoing development of the models that predict the dependence of these
missions' science yield on exozodiacal dust levels \citep{stark2019}.

The survey results in Tables~\ref{tab_measurements} and~\ref{tab_rates} show that only ~25\% of stars are dusty enough to detect with LBTI
(median 3\,$\sigma$ sensitivity of 69\,zodis for early-type stars and 144\,zodis for sun-like stars).  At these levels most stars are not
very dusty.  The median dust level is inferred from the most-likely luminosity function consistent with the HOSTS dataset of the Sun-like
stars subsample to be 3\,zodis.  From the distribution of luminosity functions that produce an acceptable fit to the data, the median level
may well be below 9\,zodis and is likely below 27\,zodis which is our 95\% upper limit.  Furthermore, almost all the HZ exozodi detections
occur in systems where cold exo-Kuiper Belt dust has been previously detected by Spitzer or Herschel; indeed, at 21\% the independently
determined frequency of cold dust in nearby stars is comparable \citep{montesinos2016} to the incidence of HZ dust at LBTI's sensitivity
level.  The HOSTS results show that the presence of detectable cold dust is usually a signpost of significant amounts of warm dust in the HZ. 
The high backgrounds indicated in these systems make them problematic targets for rocky planet spectroscopy, as the integration times
needed to characterize  atmospheres against these backgrounds are likely to be prohibitive.  High zodi levels have also been detected for a
number of stars without known cold dust, showing that the correlation is not always reliable.  However, the majority of Sun-like stars
without cold dust should have zodi levels lower than the values inferred for the full sample, and thus be more favorable targets.  

The brightness of our solar system's zodiacal light is our reference point for estimating the exozodiacal backgrounds that will affect
reflected light imaging of HZ rocky planets.  It is observed to vary with ecliptic latitude around the sky, and also with the ecliptic
longitude offset from the Sun \citep[cf. Table 9.4 of ][]{ryon2019}.  We adopt a reference line of sight through our local zodiacal
background corresponding to an ecliptic latitude of 30 degrees (the median value for targets randomly distributed over the sky), and
ecliptic longitude offset of 90 degrees (corresponding to an exoplanet target seen at maximum elongation).  Along this line of sight our
local zodiacal light has a $V$~band surface brightness of $V = 22.7$\,mag/arcsec$^2$ looking outward from the Earth.  As an external
observer's line of sight traverses both inward and outward paths through an optically thin exozodiacal cloud, it is necessary to double the
surface brightness relative to the our local observed values.  We therefore adopt a correspondence of $V = 22.0$\,mag/arcsec$^2$ to one zodi
of exozodiacal light, scaling this accordingly as we consider the effect of exozodi level on integration times for spectroscopy of HZ rocky
planets.  At $R$~band where detections of the the $0.76\,\mu$m O$_2$ feature will be sought, one zodi of exozodiacal light corresponds to
21.4\,mag/arcsec$^2$.  

The three exoplanet direct imaging missions currently under consideration would be built around 2.4\,m, 4.0\,m, or 8.0/15.0\,m telescope
apertures respectively.  Because of their different telescope sizes, each mission could tolerate different amounts of exozodi around a
fiducial target star.  We first discuss the impact of the best-fit median zodi level from HOSTS, before we discuss the implications of
assuming more conservatively zodi levels at our 1\,$\sigma$ and 95\% upper confidence limits.  Following the approach of
\citet{roberge2012}\footnote{Note that the LBTI nulling measurements were made in $N$~band and converted to units of zodis using the
approach described in Sect.~\ref{sect_null-to-zodi} with all its assumptions and limitations.  The unit of 1\,zodi is a unit of vertical
geometrical optical depth (surface density) of dust in a star's HZ.  It thus does not depend on the observing wavelength.  Predicting the
visible light brightness of the dust based on its zodi level at the relevant observing wavelength is not part of the current paper, we instead
use the predictions by \citet{roberge2012} who give a surface brightness of $\approx 22$\,mag/arcsec$^2$ for a 1\,zodi disk viewed at an
inclination of 60$^\circ$.}, for a solar analog at 10 pc observed in $R$~band ($R = 4.4$) at quadrature, the signal from 3\,zodis of dust
will exceed that of an Earth analog by factors of 42, 15, 5.4, and 1.4 for the WFIRST Starshade Rendezvous, HabEx, LUVOIR 8\,m (6.7\,m
inscribed circle), and LUVOIR 15\,m (13.5\,m inscribed circle) apertures respectively.  For this target, spectra of the $0.76\,\mu$m O$_2$
feature could be obtained against these backgrounds with
continuum $S/N \geq 10$ in reasonable integration times ($< 60$\,days), by HabEX and the two LUVOIR apertures at spectral resolution $R=140$. 
However, the median exozodi level inferred by our study would not allow the WFIRST Starshade Rendezvous mission to perform $R=50$ spectroscopy
for this target in reasonable integration times\footnote{System spectroscopy throughputs of 0.025, 0.18, 0.09, and 0.08 were adopted for the
2.4\,m, 4.0\,m, 8.0\,m, and 15.0\,m apertures respectively.}.

However this is not the end of the story.  As their apertures increase in size, each mission concept aspires to survey a larger and
progressively fainter set of targets.  The median brightnesses of their target stars are $V = 3.5$, 4.6, 5.4, and 5.7 for the Starshade
Rendezvous, HabEx, LUVOIR 8\,m, and LUVOIR 15\,m apertures respectively.  The WFIRST Starshade Rendezvous mission is capable of making the
above O$_2$ $0.76\,\mu$m spectral measurement for its median target with the median exozodi level found by the HOSTS survey, as are all three
of the other concepts.  This is a key result of the LBTI exozodi efforts: exozodi levels appear to be low enough that all of the current
mission concepts for imaging HZ rocky planets could achieve their spectral characterization objectives for their median sample target.

While the median exozodi level found by HOSTS is enabling for future missions, the formal uncertainty in the median remains a cause for
concern.  The two LUVOIR apertures and the HabEx aperture can still achieve continuum $S/N \geq 10$ for O$_2$ detection on their median
targets with the +1\,$\sigma$ HOSTS exozodi level of 9\,zodis, in less than 60\,days of integration.  For WFIRST's 2.4m aperture and
$R = 50$, this could be achieved only by relaxing the target $S/N$ to $\sim$8.  For the upper limit to the median exozodi at 95\% confidence
(27\,zodis), the achievable spectroscopic S/N on the median sample target falls below 10 for the the 4.0\,m aperture and down to 3 for the
2.4/m aperture, making it doubtful that they could achieve their mission objectives to spectrally characterize the atmospheres of habitable
zone rocky planets.  In summary, the remaining uncertainty in exozodi level poses a significant risk to the quality of the spectra that could
be obtained with apertures $\leq 4$\,m.  
 
It should be kept in mind that exozodi levels are expected to vary with each individual target.  Earth analogs could still be detected and
well-characterized even with the smaller apertures, if they were present around the nearest stars, or around stars with dust levels below the
median of the distribution.  When the dust signal is much brighter than the planet, clumps and asymmetries in the dust distribution can
become a source of confusion for exoplanet detection.  For the 4\,m aperture chosen by HabEx, \citet{defrere2012b} found that this confusion
becomes acute above the 20\,zodi level, approximately HOSTS' 95\% confidence upper limit to the median exozodi for sun-like stars.
Multi-epoch imaging could be used to distinguish between the exoplanets and exozodi clumps, as they are expected to have very different phase
functions.

\subsection{Path for further improvements}
\label{sect_sensitivity}
Currently, the main limitations of the LBTI's nulling interferometric sensitivity are of systematic nature, related to limitations of
background and low frequency detector noise removal.  The current detector of NOMIC is a Raytheon 1024$\times$1024 Si:As IBC Aquarius array
which is affected by excess low frequency noise (ELFN, \citealt{hoffmann2014}).  We are currently evaluating the possibility to upgrade NOMIC
with a new H1RG HgCdTe detector with a sensitivity cutoff at a wavelength 13\,$\mu$m.  This detector promises twice the quantum efficiency of
our current detector and not to be affected by ELFN.

In addition, telescope vibrations have been shown to limit our ability to stabilize the optical path delay (OPD) between the two primary
apertures.  Reducing the power of the strongest vibration (12\,Hz, attributed to wind induced secondary mirror swing arm vibrations) to a
level observed during the better half of the HOSTS data acquisition can reduce the statistical uncertainties of our nulling observations by
25\% to 50\% by improving the null depth of the LBTI.  This may be achieved by additional dampening of the vibrations and compensation by
more aggressive use of the OPD and Vibration Monitoring System (OVMS, \citealt{boehm2016}).  Furthermore, a larger setpoint dither pattern
\citep{ertel2018a} than used for the past HOSTS observations has recently been shown to help achieve a higher accuracy of the NSC by more
effectively breaking the degeneracy between imperfect set point and actual astrophysical null signal.

When all these improvements are implemented, the uncertainties of our null measurements will be reduced by a factor of two to three.  This
will enable us to further improve our constraints on the median zodi level and the exozodi luminosity function around future exo-Earth
imaging mission targets through a revived HOSTS survey.  Assuming our median zodi level remains unchanged by the new measurements, this will
test at a 3\,$\sigma$ confidence level whether all mission concepts discussed in Sect.~\ref{sect_imp_exoearths} will be able to achieve their
spectral characterization goal.  If the measured median zodi level changes within our current uncertainties, this could be a deciding factor
of which mission should move forward to be able to successfully detect and characterize rocky HZ planets.

In addition, there are open questions about the origin and properties of exozodiacal dust that can be answered by complementary observations
at other wavelengths from the visible to mid-infrared range \citep{mennesson2019a, gaspar2019}.  Precision interferometric observations in
the near and mid-infrared can provide constraints on the connection between HZ dust and hotter dust closer in which is critical to create a
more comprehensive picture of the dust distribution and evolution in the inner regions of planetary systems \citep{kirchschlager2017,
ertel2018b}.  Scattered light observations in the visible \citep{mennesson2019b} can constrain the dust properties and help make a connection
between the dust's infrared thermal emission and its scattered light brightness which is critical for future exo-Earth imaging missions. 
Spectrointerferometry in the LBTI's Fizeau mode \citep{spalding2018, spalding2019} provides another possibility to constrain the dust
properties and thus to better predict its brightness at different wavelengths and to learn about its origin and evolution (cometary origin,
PR drag, or local production through equilibrium or episodic/catastrophic collisions).

\section{Conclusions}
\label{sect_conc}

The HOSTS survey has been completed successfully after observing 38 stars with a median 3\,$\sigma$ sensitivity in $N$~band of
69\,zodis for early type stars and 144\,zodis for Sun-like stars.  In this paper we have presented and statistically analyzed the final
astrophysical null and zodi measurements.

We have detected significant excess around ten stars and have derived basic detection statistics with respect to other system parameters.
Almost all stars with known debris disks also show excess in our observations with derived HZ dust levels one to three orders of magnitude
higher than in our Solar system.  This correlation suggests an origin of the HZ dust in the outer disk.  It seems plausible that the two
stars with outer debris disk but without a HOSTS detection ($\sigma$\,Boo and $\tau$\,Cet) also have high HZ dust levels but that these are
too faint to be detected by our observations with weak upper limits of 140\,zodis and 120\,zodi, respectively.  However, we also found strong
detections of HZ dust around stars without a known debris disk which suggests that an alternative scenario for creating this dust may be at
play in these systems or that even tenuous cold debris disks that remain undetected by current observations may be a significant source of HZ
dust.

After accounting for sensitivity biases in our data, we found no signs of stellar spectral type or age dependence of the occurrence rates of
HZ dust in our data.  Although our small number statistics prevent us from detecting small trends, there seems to be no reason to avoid young
or early type stars for exo-Earth imaging missions due to their expected HZ dust content, except insofar as these are more likely to
have bright cold debris belts that are an indicator of high HZ dust content.  The fact that we detected bright HZ dust disks around Gyr old
stars suggests that these originated either from a recent, stochastic event, or in slowly decaying outer, Kuiper belt-like debris disks
rather than more rapidly decaying asteroid belt-like disks.

We hypothesized that at least two different types of HZ dust systems may exist; `docile' systems with moderate amounts of dust are likely
explained by a continuous delivery of dust to the HZ, while more extreme systems with large amounts of dust are likely better explained by a
catastrophic or at least an episodic dust production or delivery mechanism, or a very specific planetary system architecture that may
support a high rate of cometary influx.  Cometary delivery can contribute to both as a steady flow of comets can be present over a Gyr time
span or caused by an episodic event like a late heavy bombardment \citep{gomes2005}.  For Sun-like stars we may typically be only sensitive
enough to detect the latter.  Our statistical results can be used to validate future models of the origin and properties of exozodiacal dust. 
In addition, detailed studies of the detected exozodis will improve our understanding of their architectures and the dust production/delivery
mechanisms at play.  The combination of an improved understanding of the dust production and delivery in individual systems with population
synthesis models calibrated against our detection statistics will improve our predictive power of HZ dust levels for systems that could not
be observed.

Fitting a free form luminosity function to our zodi measurements of Sun-like stars, we derived a median zodi level of
$m = 3^{+6}_{-3}$\,zodis (95\% confidence upper limit: 27\,zodis).  Our median zodi level would suggest that all currently studied exo-Earth
imaging mission concepts will be able to achieve their mission objectives to detect and spectroscopically characterize rocky, HZ planets.  
However, more precise constraints are still required, in articular for the spectroscopic characterization of the detected planets by
missions with a primary aperture $\leq$4\,m.  We have outlined a path forward to further improve our constraints by moderate instrument
upgrades to the LBTI and a revived HOSTS survey.

We find that stars with detected, cold debris disks almost certainly have high HZ dust levels and should be avoided by future exo-Earth
imaging missions.  We find no indication that young or early type stars have higher zodi levels than old late type stars, but our limited
sample size prevents us from detecting weak correlations.

The best-fit median HZ dust level derived from our data is only a factor of a few larger than in our Solar system and consistent with it
within our 1\,$\sigma$ uncertainty.  This suggests that the Solar system's HZ dust content appears typical or only slightly low compared to
other, similar stars.  However, our uncertainties still permit the typical HZ dust levels around comparable stars to be over an order of
magnitude higher than in the Solar system.

Despite the successful completion of the HOSTS survey, there are several open questions that need to be answered in the future, specifically
with new, more sensitive LBTI observations.  The diversity of exozodi systems needs to be better understood by follow-up observations and
characterization of the detected systems to better understand the origin of the dust.  One caveat of the HOSTS observations is the weak
constraints on the dust properties and thus the scattered light brightness of exozodiacal dust in the visible from the $N$~band thermal
emission observations.  Characterizing the detected systems through multi-wavelength observations with the LBTI across the $N$~band (and in
principle possible down to the $K$~band in case of hotter dust) is critical to better constrain the dust properties and to complement future
scattered light observations of our brightest targets, e.g., with WFIRST.  The prospects for follow-up observations of HOSTS detections with
the LBTI have been discussed in detail by \citet{ertel2018spie}.

\acknowledgments

The Large Binocular Telescope Interferometer is funded by the National Aeronautics and Space Administration as part of its Exoplanet
Exploration Program.  The LBT is an international collaboration among institutions in the United States, Italy, and Germany. LBT Corporation
partners are: The University of Arizona on behalf of the Arizona university system; Instituto Nazionale di Astrofisica, Italy; LBT
Beteiligungsgesellschaft, Germany, representing the Max-Planck Society, the Astrophysical Institute Potsdam, and Heidelberg University; The
Ohio State University, and The Research Corporation, on behalf of The University of Notre Dame, University of Minnesota and University of
Virginia. Part of this research was carried out at the Jet Propulsion Laboratory, California Institute of Technology, under a contract
with the National Aeronautics and Space Administration.  GMK is supported by the Royal Society as a Royal Society University Research Fellow. 
KMM's work is supported by the NASA Exoplanets Research Program (XRP) by cooperative agreement NNX16AD44G.  This research has made extensive
use of the SIMBAD database \citep{wenger2000} and the VizieR catalogue access tool \citep{ochsenbein2000}, both operated at CDS, Strasbourg,
France, of Python, including the NumPy, SciPy, Matplotlib \citep{hunter2007}, and Astropy \citep{astropy2013} libraries, and of NASA's
Astrophysics Data System Bibliographic Services.

\setcounter{table}{0}
\begin{longrotatetable}
\begin{deluxetable*}{cccccccccccccc}
\tablecaption{Observed targets and relevant stellar parameters\label{tab_stars}}
\tablecolumns{13}
\tabletypesize{\footnotesize}
\tablehead{
HD     & Name            & Spectral & $V$   & $K$   & $N'^{~a}$     & $d$  & Age   & EEID$^b$ & fIR/nIR & Excess       & \#      & PA range$^d$ & Dates observed$^e$ \\
number &                 & Type     & (mag) & (mag) & (Jy)     & (pc) & (Myr) & (mas)    & excess  & ~references~ & SCI$^c$ & (deg)        & 
}
\startdata
\multicolumn{13}{l}{Sensitivity driven sample (Spectral types A to F5)$^f$:} \\
\hline
33111  & $\beta$\,Eri    &  A3\,IV  & 2.782 & 2.38  & 3.7  & 27.4 &  761 & 248   &   N/N   & 1,2,3     &   2    & [22, 37]                & 2017 Feb 10 \\[2pt]
38678  & $\zeta$\,Lep    & A2\,IV-V & 3.536 & 3.31  & 2.1  & 21.6 &  587 & 176   &   Y/Y   & 4,5       &  1/6   & [6, 8]                  & 2017 Dec 23 \\[2pt]
81937  & 23\,UMa         &  F0\,IV  & 3.644 & 2.73  & 2.6  & 23.8 & 1172 & 168   &   N/--  & 6         &   1    & [-158, 175]$^\text{N}$  & 2016 Nov 15 \\[-3pt]
       &                 &          &       &       &      &      &      &       &         &           &   2    & [-124, -144]$^\text{N}$ & 2017 Feb 11 \\[-3pt]
       &                 &          &       &       &      &      &      &       &         &           &   2    & [173, 159]$^\text{N}$   & 2018 Mar 30 \\[2pt]
95418  & $\beta$\,UMa    &  A1\,IV  & 2.341 & 2.38  & 4.2  & 24.5 &  404 & 316   &   Y/N   & 5,7       &   4    & [-145, 168]$^\text{N}$  & 2017 Apr 03 \\[2pt]
97603  & $\delta$\,Leo   &  A5\,IV  & 2.549 & 2.26  & 3.9  & 17.9 &  718 & 278   &   N/N   & 1,2,5     &   2    & [-54, -46]              & 2017 Feb 10 \\[-3pt]
       &                 &          &       &       &      &      &      &       &         &           &   2    & [21, 52]                & 2017 May 12 \\[2pt]
102647 & $\beta$\,Leo    &  A3\,V   & 2.121 & 1.92  & 6.9  & 11.0 &   45 & 336   &   Y/Y   & 5,7       &   2    & [41, 58]                & 2015 Feb 08 \\[2pt]
103287 & $\gamma$\,UMa   &  A0\,IV  & 2.418 & 2.43  & 3.7  & 25.5 &  334 & 308   &   N/--  & 1,2,7     &   2    & [-163, 168]$^\text{N}$  & 2017 Apr 06 \\[-3pt]
       &                 &          &       &       &      &      &      &       &         &           &   2    & [128, 113]$^\text{N}$   & 2017 May 01 \\[2pt]
106591 & $\delta$\,UMa   &  A2\,V   & 3.295 & 3.10  & 2.0  & 24.7 &  234 & 199   &   N/N   & 1,2,5     &   2    & [-113, 167]$^\text{N}$  & 2017 Feb 09 \\[-3pt]
       &                 &          &       &       &      &      &      &       &         &           &   2    & [150, 133]$^\text{N}$   & 2017 May 21 \\[-3pt]
       &                 &          &       &       &      &      &      &       &         &           &   3    & [171, 118]$^\text{N}$   & 2018 May 25 \\[2pt]
108767 & $\delta$\,Crv   &  A0\,IV  & 2.953 & 3.05  & 2.3  & 26.6 &  175 & 251   &   N/Y   & 1,2,3     &   2    & [-20, -7]               & 2017 Feb 10 \\[2pt]
109085 & $\eta$\,Crv     &  F2\,V   & 4.302 & 3.54  & 1.8  & 18.3 & 1433 & 125   &   Y/N   & 8,9       &   3    & [-5, 32]                & 2014 Feb 12 \\[2pt]
128167 & $\sigma$\,Boo   &  F4\,V   & 4.467 & 3.47  & 1.4  & 15.8 & 1703 & 117   &   Y$^g$/N & 1,5     &   1    & [-50, 70]               & 2017 Apr 03 \\[-3pt]
       &                 &          &       &       &      &      &      &       &         &           &   2    & [-74, -66]              & 2017 Apr 06 \\[2pt]
129502 & $\mu$\,Vir      &  F2\,V   & 3.865 & 2.89  & 2.6  & 18.3 & 1753 & 151   &   N/N   & 1,3       &   3    & [-26, 4]                & 2017 Feb 10 \\[2pt]
172167 & $\alpha$\,Lyr   &  A0\,V   & 0.074 & 0.01  & 38.6 & 7.68 &  428 & 916   &   Y/Y   & 5,7       &   2    & [-106, -125]$^\text{N}$ & 2017 Apr 06 \\[-3pt] 
       &                 &          &       &       &      &      &      &       &         &           &   2    & [-89, -100]$^\text{N}$  & 2018 Mar 28 \\[2pt]
187642 & $\alpha$\,Aql   &  A7\,V   & 0.866 & 0.22  & 21.6 & 5.13 &  739 & 570   &   N/Y   & 1,2,5,10  &   2    & [-52, 20]$^\text{N}$    & 2017 May 12 \\[2pt]
203280 & $\alpha$\,Cep   &  A8\,V   & 2.456 & 1.85  & 7.0  & 15.0 &  958 & 294   &   N/Y   & 1,2,5,10  &   1    & [130, 121]$^\text{N}$   & 2016 Oct 16 \\[2pt]
\hline
\multicolumn{13}{l}{Sun-like stars sample (Spectral types F6 to K8)$^f$:} \\
\hline
9826   & $\upsilon$\,And &  F9\,V   & 4.093 & 2.84  & 2.4  & 13.5 & 4000 & 136   &   N/N   & 5,11      &   2    & [-118, 158]$^\text{N}$  & 2017 Dec 20 \\[2pt]
10476  & 107\,Psc        &  K1\,V   & 5.235 & 3.29  & 2.0  & 7.53 & 4990 &  90   &   N/N   & 1,5,11,12 &   1    & [-20, 12]               & 2016 Nov 14 \\[-3pt]
       &                 &          &       &       &      &      &      &       &         &           &   2    & [-40, 22]               & 2016 Nov 16 \\[2pt]
10700  & $\tau$\,Cet     &  G8\,V   & 3.489 & 1.68  & 5.4  & 3.65 & 5800 & 182   &   Y/Y   & 5,13      &   2    & [5, 29]                 & 2018 Jan 05 \\[2pt]
16160  & GJ\,105\,A      &  K3\,V   & 5.815 & 3.45  & 1.5  & 7.18 & 6100 &  73   &   N/--  & 1,11,12   &   1    & [9, 19]                 & 2016 Nov 15 \\[2pt]
22049  & $\epsilon$\,Eri &  K2\,V   & 3.721 & 1.67  & 7.4  & 3.22 &  600 & 172   &   Y/N   & 8,14      &   2    & [-4, 16]                & 2017 Dec 20 \\[-3pt]
       &                 &          &       &       &      &      &      &       &         &           &   2    & [-19, 4]                & 2017 Dec 23 \\[2pt]
30652  & 1\,Ori          &  F6\,V   & 3.183 & 2.08  & 4.8  & 8.07 & 1200 & 205   &   N/N   & 1,5,11,12 &   2    & [0, 25]                 & 2017 Feb 09 \\[-3pt]
       &                 &          &       &       &      &      &      &       &         &           &   2    & [5, 23]                 & 2017 Dec 20 \\[2pt]
34411  & $\lambda$\,Aur  &  G1\,V   & 4.684 & 3.27  & 1.8  & 12.6 & 7700 & 105   &   N/--  & 11,15     &   2    & [101, 83]  N            & 2017 Jan 29 \\[2pt]
48737  & $\xi$\,Gem      & F5\,IV-V & 3.336 & 2.13  & 4.3  & 18.0 & 2000 & 196   &   --/N  & 5         &   2    & [0, 19]                 & 2016 Nov 14 \\[-3pt]
       &                 &          &       &       &      &      &      &       &         &           &   1    & [-44, 21]               & 2016 Nov 15 \\[2pt]
78154  & 13\,UMa         &  F7\,V   & 4.809 & 3.53  & 1.2  & 20.4 & 4900 &  99   &   N/--  & 1         &   2    & [-168, 163]  N          & 2018 Mar 29 \\[-3pt]
       &                 &          &       &       &      &      &      &       &         &           &   1    & [141, 127]  N           & 2018 Mar 30 \\[2pt]
88230  & GJ\,380         &  K8\,V   & 6.598 & 3.21  & 1.9  & 4.87 & 1200 &  65   &   N$^h$/-- & 16     &   2    & [-143, -167]  N         & 2017 Apr 06 \\[2pt]
89449  & 40\,Leo         & F6\,IV-V & 4.777 & 3.65  & 1.1  & 21.4 & 3100 &  98   &   N/--  & 1,6       &   2    & [-58, -16]  N           & 2017 Feb 09 \\[2pt]
102870 & $\beta$\,Vir    &  F9\,V   & 3.589 & 2.31  & 4.3  & 10.9 & 4400 & 173   &   N/N   & 5,11      &   2    & [-25, -3]               & 2017 Dec 20 \\[-3pt]
       &                 &          &       &       &      &      &      &       &         &           &   2    & [13, 26]                & 2018 Mar 30 \\[2pt]
120136 & $\tau$\,Boo     &  F6\,IV  & 4.480 & 3.36  & 1.7  & 15.6 & 1300 & 114   &   N/N   & 3,11,15   &   2    & [28, 57]                & 2017 May 12 \\[-3pt]
       &                 &          &       &       &      &      &      &       &         &           &   2    & [10, 38]                & 2018 Mar 30 \\[2pt]
126660 & $\theta$\,Boo   &  F7\,V   & 4.040 & 2.81  & 3.1  & 14.5 &  500 & 147   &   N$^i$/--  & 1,11,12 & 1    & [-170, 164]  N          & 2017 Feb 09 \\[-3pt]
       &                 &          &       &       &      &      &      &       &         &           &   2    & [-170, 152]  N          & 2017 Apr 11 \\[-3pt]
       &                 &          &       &       &      &      &      &       &         &           &   2    & [-158, 176]  N          & 2018 May 23 \\[2pt]
141004 & $\lambda$\,Ser  & G0\,IV-V & 4.413 & 2.98  & 2.4  & 12.1 & 5300 & 121   &   N/N   & 1,5,12,17 &   2    & [7, 24]                 & 2017 May 01 \\[2pt]
142373 & $\chi$\,Her     &  G0\,V   & 4.605 & 3.12  & 2.0  & 15.9 & 6210 & 111   &   N/N   & 1,5,6,12  &   3    & [131, 99]  N            & 2017 Apr 11 \\[2pt]
142860 & $\gamma$\,Ser   &  F6\,IV  & 3.828 & 2.63  & 2.9  & 11.3 & 4600 & 151   &   N/N   & 1,5,12,15 &   2    & [-29, -8]               & 2017 Apr 06 \\[-3pt]
       &                 &          &       &       &      &      &      &       &         &           &   2    & [-30, 12]               & 2017 May 21 \\[2pt]
157214 & 72\,Her         &  G0\,V   & 5.381 & 3.84  & 1.0  & 14.3 & 6900 &  79   &   N/--  & 11        &   2    & [-84, -86]              & 2018 May 23 \\[-3pt]
       &                 &          &       &       &      &      &      &       &         &           &   2    & [-84, -85]              & 2018 May 25 \\[2pt]
173667 & 110\,Her        &  F6\,V   & 4.202 & 3.03  & 2.2  & 19.2 & 2200 & 131   &   Y$^j$/Y & 5,11,16 &   2    & [-56, -31]              & 2017 Apr 08 \\[-3pt]
       &                 &          &       &       &      &      &      &       &         &           &   3    & [-63, -50]              & 2018 Mar 30 \\[2pt]
185144 & $\sigma$\,Dra   &  G9\,V   & 4.664 & 2.83  & 2.7  & 5.76 & 3500 & 113   &   N/N   & 5,11,15   &   2    & [-143, -163]  N         & 2017 May 01 \\[2pt]
201091 & 61\,Cyg\,A      &  K5\,V   & 5.195 & 2.36  & 4.4  & 3.49 & 7000 & 106   &   N/N   & 5,13      &   2    & [-92, -115]  N          & 2018 May 23 \\[-3pt]
       &                 &          &       &       &      &      &      &       &         &           &   2    & [-92, -100]  N          & 2018 May 25 \\[2pt]
215648 & $\xi$\,Peg\,A   &  F6\,V   & 4.203 & 2.90  & 2.2  & 16.3 & 5000 & 132   &   N/N   & 1,6,12    &   1    & [20, 30]                & 2016 Nov 14 \\[-3pt]
       &                 &          &       &       &      &      &      &       &         &           &   2    & [4, 25]                 & 2016 Nov 16 \\[2pt]
222368 & $\iota$\,Psc    &  F7\,V   & 4.126 & 2.80  & 2.4  & 13.7 & 5000 & 137   &   N/--  & 6         &   2    & [-33, 37]               & 2017 Nov 10 \\[2pt]
\hline
\enddata
\tablecomments{Magnitudes are given in the Vega system.\\
$^a$~Predicted flux in NOMIC $N'$~filter.
$^b$~Earth Equivalent Insolation Distance \citep{weinberger2015}.
$^c$~Number of calibrated science pointings obtained.
$^d$~Approximate parallactic angle (PA) range covered by the observations.  In practice the PA coverage in this range is not uniform due to
changing sky rotation and observations being unevenly distributed in time.  Northern targets are marked by a `N' as they culminate at the
$-180^\circ \equiv 180^\circ$ discontinuity of the PA rather than 0$^\circ$.
$^e$~All data including auxiliary information such as weather conditions, exact observing time, and hour angle coverage, are available in the
HOSTS archive (http://lbti.ipac.caltech.edu/).
$^f$~Sect.~\ref{sect_obs}.
$^g$~Mis-classified by \citet{gaspar2013} as no excess.
$^h$~Cold excess \citep{eiroa2013} likely background contamination \citep{gaspar2014}.
$^i$~Tentative detection at 2.5\,$\sigma$ \citep{montesinos2016}, may have a faint excess.
$^j$~Marginal excesses from Spitzer at 70\,$\mu$m \citep{trilling2008} and Herschel at 70\,$\mu$m and 100\,$\mu$m excesses \citep{eiroa2013, marshall2013}, taken together likely a faint excess.
\\
References are:
Spectral type: SIMBAD;
$V$~magnitude: \citet{kharchenko2007};
$K$~magnitude: \citet{gezari1993} and the Lausanne photometric data base (http://obswww.unige.ch/gcpd/gcpd.html);
$N$~band flux and EEID: \citet{weinberger2015};
Distance: \citet{vanleeuwen2007};
Excess: (1)~\citet{gaspar2013}, (2)~\citet{thureau2014}, (3)~\citet{ertel2014b}, (4)~\citet{mannings1998}, (5)~\citet{absil2013},
(6)~\citet{beichman2006}, (7)~\citet{su2006}, (8)~\citet{absil2006}, (9)~\citet{aumann1988}, (10)~\citet{rieke2005},
(11)~\citet{trilling2008}, (12)~\citet{montesinos2016}, (13)~\citet{greaves2004}, (14)~\citet{aumann1985}, (15)~\citet{lawler2009}, 
(16)~\citet{eiroa2013}, (17)~\citet{koerner2010}.}
\end{deluxetable*}
\end{longrotatetable}

\begin{longrotatetable}
\begin{deluxetable*}{|cc||cc|cc|ccc|ccccc|}
\tablecaption{Basic stellar parameters, astrophysical null measurements, and zodi levels for the stars observed by the HOSTS survey
\label{tab_measurements}}
\tablecolumns{16}
\tablewidth{0pt}
\tablehead{
\hline
\multicolumn{2}{|c||}{Aperture $\rightarrow$} & \multicolumn{2}{c|}{8\,pix} & \multicolumn{2}{c|}{13\,pix} & \multicolumn{3}{c|}{conservative} &           &                              &         &            &              \\
\hline
HD      & Name           & $N_{\text{as}}$  & $\sigma_N$  & $N_{\text{as}}$  & $\sigma_N$  & $r_{\text{ap}}$  & $N_{\text{as}}$  & $\sigma_N$  & aperture  & $N_{\text{as,1}}$            & $z$     & $\sigma_z$ & $z/\sigma_z$ \\
number  &                & (\%)             & (\%)        & (\%)             & (\%)        & (pix)            & (\%)             & (\%)        & for zodi  & (\%)                         & (zodi)  & (zodi)     &
}
\startdata
\multicolumn{14}{l}{Sensitivity driven sample (Spectral types A to F5):} \\
\hline
33111  & $\beta$\,Eri    &  -0.004          &    0.110    &   0.168          &   0.119     &  18              &   0.372          &  0.176      &  13\,pix  & 5.27$\times$10$^{\text{-3}}$ &   31.9  &   22.6     &   1.4  \\
38678  & $\zeta$\,Lep    &   1.795          &    0.205    &   1.609          &   0.313     &  25              &   3.496          &  0.214      &  cons.    & 5.77$\times$10$^{\text{-3}}$ &  605.8  &   37.2     &  16.3  \\
81937  & 23\,UMa         &  -0.065          &    0.061    &  -0.032          &   0.078     &  25              &  -0.135          &  0.142      &  13\,pix  & 3.31$\times$10$^{\text{-3}}$ &   -9.8  &   23.5     &  -0.4  \\
95418  & $\beta$\,UMa    &   0.920          &    0.055    &   1.019          &   0.060     &  33              &   1.655          &  0.102      &  13\,pix  & 7.45$\times$10$^{\text{-3}}$ &  136.7  &    8.0     &  17.1  \\
97603  & $\delta$\,Leo   &   0.028          &    0.051    &   0.033          &   0.055     &  32              &  -0.013          &  0.143      &  13\,pix  & 6.10$\times$10$^{\text{-3}}$ &    5.5  &    9.0     &   0.6  \\
102647 & $\beta$\,Leo    &   0.470          &    0.050    &   0.420          &   0.054     &  32              &   1.160          &  0.333      &  8\,pix   & 4.54$\times$10$^{\text{-3}}$ &  103.5  &   11.0     &   9.4  \\
103287 & $\gamma$\,UMa   &  -0.037          &    0.033    &   0.003          &   0.031     &  34              &   0.083          &  0.080      &  13\,pix  & 8.00$\times$10$^{\text{-3}}$ &    0.4  &    3.9     &   0.1  \\
106591 & $\delta$\,UMa   &   0.453          &    0.065    &   0.503          &   0.082     &  28              &   0.924          &  0.144      &  8\,pix   & 5.38$\times$10$^{\text{-3}}$ &   84.2  &   12.1     &   7.0  \\
108767 & $\delta$\,Crv   &  -0.333          &    0.131    &  -0.243          &   0.199     &  26              &   0.933          &  0.365      &  13\,pix  & 9.01$\times$10$^{\text{-3}}$ &  -26.9  &   22.1     &  -1.2  \\
109085 & $\eta$\,Crv     &   4.410          &    0.350    &   4.580          &   0.460     &  24              &   4.710          &  0.890      &  8\,pix   & 2.26$\times$10$^{\text{-3}}$ & 1952.3  &  154.9     &  12.6  \\
128167 & $\sigma$\,Boo   &  -0.019          &    0.096    &  -0.006          &   0.118     &  22              &   0.417          &  0.252      &  13\,pix  & 2.46$\times$10$^{\text{-3}}$ &   -2.3  &   48.0     &   0.0  \\
129502 & $\mu$\,Vir      &  -0.006          &    0.092    &   0.183          &   0.110     &  25              &   0.192          &  0.198      &  13\,pix  & 2.64$\times$10$^{\text{-3}}$ &   69.2  &   41.8     &   1.7  \\
172167 & $\alpha$\,Lyr   &   0.055          &    0.034    &   0.123          &   0.038     &  37              &   0.392          &  0.089      &  cons.    & 1.18$\times$10$^{\text{-2}}$ &   33.2  &    7.5     &   4.4  \\
187642 & $\alpha$\,Aql   &  -0.032          &    0.166    &   0.217          &   0.192     &  47              &  -0.995          &  0.356      &  13\,pix  & 4.16$\times$10$^{\text{-3}}$ &   52.1  &   46.3     &   1.1  \\
203280 & $\alpha$\,Cep   &  -0.301          &    0.376    &  -0.233          &   0.182     &  18              &  -0.075          &  0.266      &  13\,pix  & 3.91$\times$10$^{\text{-3}}$ &  -59.6  &   46.6     &  -1.3  \\
\hline
\multicolumn{14}{l}{Sun-like stars sample (Spectral types F6 to K8):} \\
\hline
9826   & $\upsilon$\,And &  -0.245          &    0.079    &  -0.287          &   0.090     &  24              &  -0.276          &  0.169      &  13\,pix  & 2.20$\times$10$^{\text{-3}}$ & -130.3  &   40.8     &  -3.2  \\
10476  & 107\,Psc        &  -0.028          &    0.083    &  -0.027          &   0.122     &  21              &   0.154          &  0.181      &  13\,pix  & 9.45$\times$10$^{\text{-4}}$ &  -28.3  &  129.4     &  -0.2  \\
10700  & $\tau$\,Cet     &   0.074          &    0.079    &  -0.014          &   0.084     &  27              &  -0.022          &  0.139      &  13\,pix  & 1.90$\times$10$^{\text{-3}}$ &   -7.5  &   43.8     &  -0.2  \\
16160  & GJ\,105\,A      &   0.228          &    0.232    &  -0.227          &   0.239     &  18              &   0.538          &  0.363      &  13\,pix  & 7.12$\times$10$^{\text{-4}}$ & -319.2  &  336.2     &  -0.9  \\
22049  & $\epsilon$\,Eri &   0.144          &    0.068    &   0.240          &   0.066     &  27              &   0.463          &  0.087      &  cons.    & 1.56$\times$10$^{\text{-3}}$ &  296.6  &   55.6     &   5.3  \\
30652  & 1\,Ori          &   0.078          &    0.098    &   0.107          &   0.101     &  28              &   0.016          &  0.165      &  13\,pix  & 2.78$\times$10$^{\text{-3}}$ &   38.4  &   36.2     &   1.1  \\
34411  & $\lambda$\,Aur  &  -0.210          &    0.095    &  -0.108          &   0.079     &  22              &   0.041          &  0.136      &  13\,pix  & 1.57$\times$10$^{\text{-3}}$ &  -69.1  &   50.6     &  -1.4  \\
48737  & $\xi$\,Gem      &   0.048          &    0.099    &   0.124          &   0.098     &  27              &   0.057          &  0.229      &  13\,pix  & 2.73$\times$10$^{\text{-3}}$ &   45.4  &   35.9     &   1.3  \\
78154  & 13\,UMa         &   0.369          &    0.102    &   0.398          &   0.144     &  22              &   0.028          &  0.181      &  13\,pix  & 1.89$\times$10$^{\text{-3}}$ &  210.1  &   75.8     &   2.8  \\
88230  & GJ\,380         &  -0.111          &    0.059    &  -0.077          &   0.056     &  20              &  -0.189          &  0.087      &  13\,pix  & 4.19$\times$10$^{\text{-4}}$ & -184.7  &  134.2     &  -1.4  \\
89449  & 40\,Leo         &   0.238          &    0.263    &  -0.018          &   0.290     &  21              &   1.278          &  0.578      &  13\,pix  & 2.19$\times$10$^{\text{-3}}$ &   -8.4  &  132.5     &  -0.1  \\
102870 & $\beta$\,Vir    &  -0.069          &    0.039    &  -0.054          &   0.049     &  26              &  -0.172          &  0.098      &  13\,pix  & 2.11$\times$10$^{\text{-3}}$ &  -25.5  &   23.0     &  -1.1  \\
120136 & $\tau$\,Boo     &   0.111          &    0.108    &  -0.112          &   0.111     &  22              &   0.300          &  0.216      &  13\,pix  & 2.09$\times$10$^{\text{-3}}$ &  -53.8  &   53.0     &  -1.0  \\
126660 & $\theta$\,Boo   &   0.280          &    0.052    &   0.329          &   0.066     &  24              &   0.441          &  0.083      &  8\,pix   & 1.89$\times$10$^{\text{-3}}$ &  148.2  &   27.7     &   5.4  \\
141004 & $\lambda$\,Ser  &   0.015          &    0.036    &   0.025          &   0.047     &  23              &  -0.107          &  0.117      &  13\,pix  & 1.68$\times$10$^{\text{-3}}$ &   15.1  &   28.1     &   0.5  \\
142373 & $\chi$\,Her     &  -0.063          &    0.052    &   0.112          &   0.061     &  22              &   0.071          &  0.083      &  13\,pix  & 1.63$\times$10$^{\text{-3}}$ &   69.1  &   37.2     &   1.9  \\
142860 & $\gamma$\,Ser   &   0.037          &    0.044    &  -0.009          &   0.058     &  25              &   0.023          &  0.079      &  13\,pix  & 2.35$\times$10$^{\text{-3}}$ &   -3.7  &   24.6     &  -0.2  \\
157214 & 72\,Her         &   0.713          &    0.146    &   0.600          &   0.173     &  20              &   0.674          &  0.193      &  8\,pix   & 1.21$\times$10$^{\text{-3}}$ &  587.5  &  120.5     &   4.9  \\
173667 & 110\,Her        &   0.152          &    0.070    &   0.194          &   0.087     &  24              &   0.621          &  0.120      &  cons.    & 2.64$\times$10$^{\text{-3}}$ &  234.9  &   45.4     &   5.2  \\
185144 & $\sigma$\,Dra   &   0.027          &    0.052    &  -0.075          &   0.071     &  22              &  -0.096          &  0.096      &  13\,pix  & 1.25$\times$10$^{\text{-3}}$ &  -60.2  &   56.9     &  -1.1  \\
201091 & 61\,Cyg\,A      &   0.060          &    0.053    &   0.047          &   0.050     &  22              &   0.126          &  0.231      &  13\,pix  & 7.00$\times$10$^{\text{-4}}$ &   66.7  &   71.3     &   0.9  \\
215648 & $\xi$\,Peg\,A   &   0.154          &    0.121    &   0.226          &   0.167     &  23              &   0.198          &  0.214      &  13\,pix  & 2.22$\times$10$^{\text{-3}}$ &  101.7  &   75.0     &   1.4  \\
222368 & $\iota$\,Psc    &  -0.099          &    0.127    &   0.016          &   0.133     &  23              &  -0.062          &  0.271      &  13\,pix  & 2.24$\times$10$^{\text{-3}}$ &    7.2  &   59.4     &   0.1  \\
\enddata
\tablecomments{Table columns are:\\
$N_{\text{as}}$ -- astrophysical null measurement in a given aperture\\
$\sigma_N$ -- uncertainty of the null measurement\\
$r_{\text{ap}}$ -- radius of the conservative aperture\\
$N_{\text{as,1}}$ -- astrophysical null expected from 1\,zodi\\
$z$ -- final zodi measurement\\
$\sigma_z$ -- uncertainty on final zodi measurement\\
$z/\sigma_z$ -- significance of zodi measurement.}
\end{deluxetable*}
\end{longrotatetable}

\facilities{LBT (LBTI/NOMIC).}

\bibliographystyle{aasjournal}
\bibliography{bibtex_new}

\appendix
\section{Modeling cookbook for LBTI null measurements}
\label{app_modeling}

We provide here a modeling cookbook for LBTI null measurements in the context of exozodiacal dust observations, intended to aid other teams
in the modeling of our data using their own tools.  A complete description of this modeling approach can be found in \citet{kennedy2015a}
with minor updates described by \cite{ertel2018a}.

\subsection{High level description of the data}

The concept of the data produced by our observations is illustrated in Fig.~\ref{fig_nulling_data}.  In the case of nulling interferometry,
the LBTI combines the light from the two apertures of the LBT in phase opposition in the pupil plane before re-imaging the target on the
detector.  Light at zero optical path delay (OPD) between the two sides (on axis or off axis perpendicular to the baseline between the two
apertures) is suppressed.  An offset on sky with a component in the direction of the interferometric baseline results in a non-zero OPD
between the two sides, so that the two light beams are out of phase.  This results in a sinusoidal transmission pattern of stripes
(Fig.~\ref{fig_nulling_data}, second column) perpendicular to the telescope baseline projected on sky with minimum transmission (dark
fringes) if the OPD is a multiple of the observing wavelengths and maximum transmission (bright fringes) half way between the dark fringes.  
The fringe pattern is always parallel to a great circle through the target and zenith, i.e., the elevation direction due to the LBT's
altitude-azimuth mount.  This transmission pattern is multiplied (Fig.~\ref{fig_nulling_data}, third column) with the angular brightness
distribution of the source on sky (Fig.~\ref{fig_nulling_data}, first column).  The beam combination in the pupil plane and re-imaging on the
detector means that the image of the source multiplied by the transmission pattern is then convolved with the single aperture telescope PSF
(Fig.~\ref{fig_nulling_data}, fourth column), which produces the final image on the detector.  Due to sky rotation during the observations,
the sky is rotating under the transmission pattern with parallactic angle.  Each of the null measurements in Table~\ref{tab_measurements} is
a combination of measurements at a range of parallactic angles.

Because the star is marginally resolved by our observations, part of the star light is transmitted through the system.  Furthermore, there is
an instrumental null leak due to imperfections of the system.  In practice, these effects are calibrated out during the data reduction and
the null measurements presented in Table~\ref{tab_measurements} are representative of the supposed circumstellar disk alone.  The
uncertainties from these calibrations are considered in the errors of the null measurements.

\subsection{Monochromatic case}
Due to the relatively large uncertainties of our measurements, it is typically sufficient to simplify the problem by considering a
monochromatic case.  The implications from our broad-band observations and the cases where chromatic effects need to be taken into account
are discussed in the next section.  Here, we provide a step-by-step guide to forward model a single nulling observation using an arbitrary
disk model in the monochromatic case:

\begin{figure}
 \centering
 \includegraphics[angle=0,width=\linewidth]{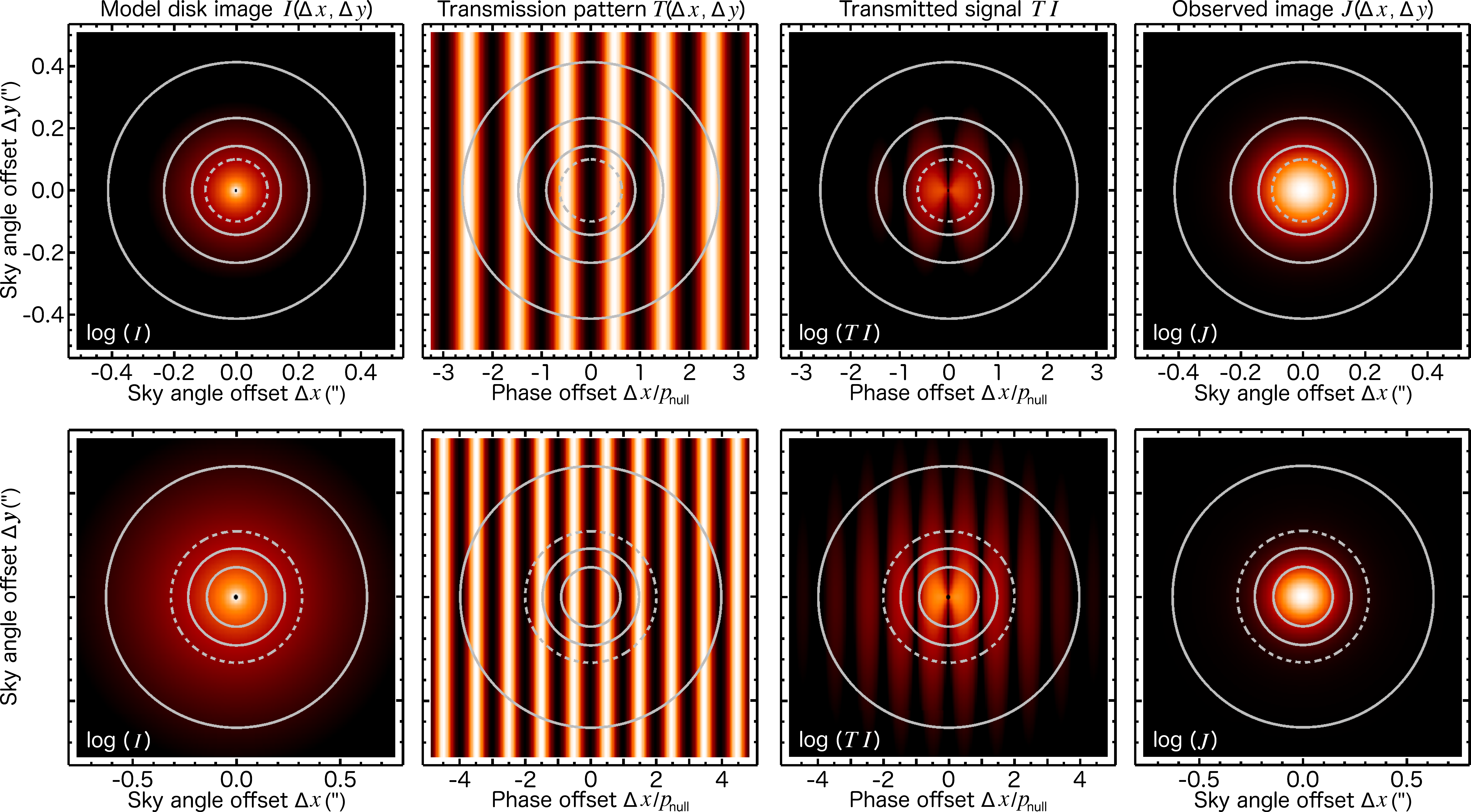}
 \caption{Illustration of the concept of LBTI nulling data and the basic steps of modeling LBTI data.  Simulations are shown for stars with
   luminosities of 1\,L$_\odot$ (\emph{top}) and 10\,L$_\odot$ (\emph{bottom}) at a distance of 10\,pc.}
 \label{fig_nulling_data}
\end{figure}

\begin{enumerate}
  \item Simulate a disk image $I\left(\Delta\alpha, \Delta\delta \right)$ from the model at 11.11\,$\mu$m in orientation North up, East left,
    where $\Delta\alpha$ and $\Delta\delta$ are angular RA and DEC sky offsets from the star, respectively.
  \item Rotate around the position of the star by the parallactic angle (PA) of the observation\footnote{The parallactic angle range of our
    observations can be found in Table~\ref{tab_stars}.  At first order, a fixed number of models (e.g., 6 for one model per nod position)
    per science pointing (\#~SCI in Table~\ref{tab_stars}) can be distributed evenly across this PA range.  In practice, individual
    observations are not evenly spaced and have varying sensitivities, and the speed of the sky rotation changes over time.  These effects
    can to some extent be neglected.  More in-depth modeling may involve downloading the data from the HOSTS archive at
    http://lbti.ipac.caltech.edu/ and to simulate observations at the exact PAs of the data as well as weighting the model points by the
    uncertainties of the individual measurements.} to obtain an image $I\left(\Delta x, \Delta y\right)$ that is in the correct sky
    orientation, where $\Delta x = x-x_0$ and $\Delta y = y-y_0$ are cartesian sky offsets from the cartesian image coordinates $x_0$ and
    $y_0$ of the star in angular units.
  \item Create an image of the transmission pattern
    $T\left(\Delta x, \Delta y \right) = sin^2 \left( \pi~\Delta x/p_{\rm{null}} \right)$, where
    $p_{\rm{null}} = \lambda/B$ is the angular period of the transmission pattern at the wavelength $\lambda = 11.11\,\mu$m and
    (fixed) interferometric baseline $B = 14.4$\,m.
  \item Create an image of the LBTI's single aperture PSF.  This can be approximated by a two-dimensional Gaussian
    $G\left(\Delta x, \Delta y\right)$ with $\textit{FWHM} = 313$\,mas (larger than the PSF of an 8.4\,m primary aperture due to an
    undersized pupil stop).
  \item A simulated LBTI image is then
    $J\left(\Delta x, \Delta y\right) = \left[I\left(\Delta x, \Delta y\right) T\left(\Delta x, \Delta y\right)\right] \circ
    G\left(\Delta x, \Delta y\right)$, where $\circ$ is the convolution operator.
  \item Perform aperture photometry on $J\left(\Delta x, \Delta y\right)$ for any of the apertures listed in Table~\ref{tab_measurements}.  
    Use a background annulus as it may include some source flux for very extended disks.  We chose the inner radius of the background annulus
    to be the radius of our conservative aperture in Table~\ref{tab_measurements} plus 17.9\,mas (one NOMIC pixel) and the outer radius was
    chosen so that the background annulus has the same area as the photometric aperture.
  \item Divide the photometric measurement of the transmitted disk flux by the flux of the star to obtain a simulated null measurement at a
    given parallactic angle.
  \item Average the individual, simulated null measurements and compare the result to the observed null values in
    Table~\ref{tab_measurements}.
\end{enumerate}

\subsection{Extension to broad-band case}
The effect of broad-band observations is that the transmission pattern is smeared out at large separations ($\gtrsim 300$\,mas) from the
star.  This effect is still negligible compared to our measurement uncertainties for smooth dust distributions.  Furthermore, observations
over a range of parallactic angles also smear the transmission pattern at large separations, which is taken into account by the modeling
approach described above.  However, for systems with a very large angular size of the HZ (very nearby stars with high luminosity such as
$\alpha$\,Lyr), if significant disk structures such as azimuthal clumps are considered, or if null measurements over a small parallactic
angle range are to be modeled individually, this effect may need to be considered.  Strong wavelength dependence of the emission, for example
in the case of a strong spectral silicate feature may also require considering the broad-band effects of our observations.

In this case, the above described approach needs to be applied to images at a range of wavelengths across the NOMIC $N'$ filter.  The
transmission of the filter, detector quantum efficiency, and the atmospheric transmission in the clear weather in which nulling observations
are usually carried out are fairly constant across the filter with cut-off wavelengths at $9.81\mu$m and 12.41$\mu$m.  The wavelength
dependence of the angular transmission pattern and the single aperture PSF (scaling linearly with $\lambda/11.11\,\mu$m) need to be taken
into account.  The simulated null measurements from all wavelengths and parallactic angles can then be averaged and compared to the
measurements in Table~\ref{tab_measurements} analogous to point~8 in the previous section.

\end{document}